\documentclass[12pt]{article}
\usepackage{amsmath}
\usepackage{amsfonts}
\usepackage{fancyhdr}
\usepackage{latexsym}
\usepackage{graphicx}
\usepackage{floatflt}
\usepackage{epsfig}
\usepackage{pstricks}
\def \fb        {{\rm \, fb}}
\def \ipb       {{\rm \, pb^{-1}}}
\def \eV        {{\rm \,  eV}}
\def \keV       {{\rm \, keV}}
\def \MeV       {{\rm \, MeV}}
\def \GeV       {{\rm \, GeV}}
\def \TeV       {{\rm \, TeV}}

\def \GeVcc     {\GeV/c^2}

\def\ga{\mathrel{\raise.3ex\hbox{$>$\kern-.75em\lower1ex\hbox{$\sim$}}}}
\def\la{\mathrel{\raise.3ex\hbox{$<$\kern-.75em\lower1ex\hbox{$\sim$}}}}
\newcommand {\lesssim}
     {\,\raisebox{-0.6ex}{$\stackrel{\textstyle<}{\textstyle\sim}$}\,}
\newcommand {\gtrsim}
     {\,\raisebox{-0.6ex}{$\stackrel{\textstyle>}{\textstyle\sim}$}\,}

\newcommand {\BR}         {{\rm BR}\,}

\newcommand {\gradi}    {^\circ}

\newcommand {\cm}         {\rm \; cm}

\newcommand {\T}          {\rm \; T}

\newcommand {\bfell}      {\ell\kern-0.4em
                           \ell\kern-0.4em
                           \ell\kern-0.4em
                           \ell }
\newcommand {\obfell}     {\overline{\ell}\kern-0.4em
                           \overline{\ell}\kern-0.4em
                           \overline{\ell}\kern-0.4em
                           \overline{\ell}}
\newcommand {\bfH}      {\; {\cal H}\kern-0.5em \kern-0.4em
                           {\cal H}\kern-0.5em \kern-0.4em
                           {\cal H}\kern0.1em }
\newcommand {\obfH}     {\; \overline{\cal H}\kern-0.5em \kern-0.4em 
                           \overline{\cal H}\kern-0.5em \kern-0.4em 
                           \overline{\cal H}\kern0.1em }
\def \b             {{\mathrm b}}

\def \charm         {{\mathrm c}}

\def \u             {{\mathrm u}}
\def \e             {{\mathrm e}}
\def \q             {{\mathrm q}}

\def \p             {{\mathrm p}}

\def \Z             {{\mathrm Z}}
\def \W             {{\mathrm W}}

\newcommand {\dM}         {\Delta M}

\newcommand {\nnc}        {{\overline{\mathrm N}_{95}}}

\newcommand {\sto}     {{\tilde{\mathrm{t}}}}

\newcommand {\glu}     {{\mathrm{\tilde{g}}}}

\newcommand {\sbot}    {{\tilde{\mathrm{b}}}}
\newcommand {\squa}    {{\tilde{\mathrm{q}}}}

\newcommand {\sqL}     {{\tilde{\mathrm{q}}_{\rm L}}}
\newcommand {\sqR}     {{\tilde{\mathrm{q}}_{\rm R}}}
\newcommand {\snu}     {{\tilde{\nu}}}

\newcommand {\neu}     {{\chi}}

\newcommand {\thstop} {\mathrm{\theta_{\tilde{t}}}}
\newcommand {\thsbot} {\mathrm{\theta_{\tilde{b}}}}
\newcommand {\thsqua} {\mathrm{\theta_{\tilde{q}}}}

\newcommand {\tanb}{\tan\beta}
\newcommand {\ee}    {{\e ^+ \e ^-}}
\newcommand {\fbody} {{\sto \to \b \chi {\rm f \bar{f}'}}}
\newcommand {\gaga}  {\gamma\gamma}
\newcommand {\ggqq}  {\gamma\gamma \rightarrow \q\overline{\q}}
\newcommand {\ggtt}  {\gamma\gamma \rightarrow \tau^{+}\tau^{-}}

\newcommand {\ththr}   {{\theta_{\rm thrust}}}

\newcommand {\acop}    {{\Phi_{\rm acop}}}
\newcommand {\acopt}   {{\Phi_{\rm acop_T}}}
\newcommand {\thpoint} {\theta_{\rm point}}
\newcommand {\thscat}  {\theta_{\rm scat}}
\newcommand {\etwelve} {E_{12\gradi}}
\newcommand {\ethirty} {E_{30\gradi}}

\newcommand {\phimiss} {{\phi_{\vec{p}_{\rm miss}}}}
\newcommand {\ewedge}  {E(\phi_{\vec{p}_{\rm miss}}\pm 15\gradi)}
\newcommand {\evis}    {E_{\rm vis}}
\newcommand {\etot}    {E_{\rm vis}}

\newcommand {\mvis}    {M_{\rm vis}}

\newcommand {\mmis}    {M_{\rm miss}}
\newcommand {\mhad}    {M^{\rm ex \, \ell_1}_{\rm vis}}

\newcommand {\ehad}    {E^{\rm NH}_{\rm vis}}

\newcommand {\nch}     {{N_{\rm ch}}}

\newcommand {\pmis}    {{\vec{p}_{\rm miss}}}
\newcommand {\thmiss}  {{\theta_{\pmis}}}
\newcommand {\pt}      {{p_{\rm t}}}
\newcommand {\ptch}    {{p_{\rm t}^{\rm ch}}}

\newcommand {\ptnoNH}  {{p_{\rm t}^{\rm ex \, NH}}}

\def \mx     {M_{\mathrm{eff}}} 

\begin{document}
\pagestyle{empty}
\begin{center} 
ORGANISATION EUROP\'EENNE POUR LA RECHERCHE NUCL\'EAIRE
(CERN) \\
Laboratoire Europ\'een pour la Physique des Particules
\end{center}
\vfill
\begin{flushright} 
  {\tt CERN-EP/2002-026}\\
  18 April 2002\\
\end{flushright} 
\vfill
\begin{center}
  \mathversion{bold}
  {\LARGE\bf  Search for Scalar Quarks in $\ee$ Collisions
    at $\sqrt{s}$ up to $209 \GeV$}
  \mathversion{normal}
  \vfill
  {\bf The ALEPH Collaboration$^{*})$}\\ 
  \vfill
  {\bf Abstract}
  \medskip
\end{center}
%
%
\small
Searches for scalar top, scalar bottom and mass-degenerate scalar quarks
are performed in the data collected by the ALEPH detector at
LEP, at centre-of-mass energies up to $209\GeV$, corresponding to an
integrated luminosity of $675\ipb$.
No evidence for the production of such particles is found in the decay 
channels $\sto\to \mathrm{c} / \mathrm{u} \, \chi$, $\sto \to \b \ell \snu$, 
$\sbot\to \b \chi$, $\squa \to \q \chi$ or in the stop four-body decay channel
$\fbody$ studied for the first time at LEP. The results of these
searches yield improved mass lower limits.
In particular, an absolute lower limit of $63\GeVcc$ is obtained for the stop 
mass, at 95\% confidence level, irrespective of the stop lifetime and decay 
branching ratios.
\vfill
\begin{center}
{\it (Submitted to Physics Letters B)}\\
\end{center}
\vfill
\begin{flushleft}
\rule{69.0mm}{0.2mm}\\
$^{*})$ See next pages for the list of authors
\end{flushleft}
\pagebreak

\pagestyle{empty}
\newpage
\small
%
%
\newlength{\saveparskip}
\newlength{\savetextheight}
\newlength{\savetopmargin}
\newlength{\savetextwidth}
\newlength{\saveoddsidemargin}
\newlength{\savetopsep}
\setlength{\saveparskip}{\parskip}
\setlength{\savetextheight}{\textheight}
\setlength{\savetopmargin}{\topmargin}
\setlength{\savetextwidth}{\textwidth}
\setlength{\saveoddsidemargin}{\oddsidemargin}
\setlength{\savetopsep}{\topsep}
%
%
\setlength{\parskip}{0.0cm}
\setlength{\textheight}{25.0cm}
\setlength{\topmargin}{-1.5cm}
\setlength{\textwidth}{16 cm}
\setlength{\oddsidemargin}{-0.0cm}
\setlength{\topsep}{1mm}
\pretolerance=10000
\centerline{\large\bf The ALEPH Collaboration}
\footnotesize
\vspace{0.5cm}
{\raggedbottom
\begin{sloppypar}
\samepage\noindent
A.~Heister,
S.~Schael
\nopagebreak
\begin{center}
\parbox{15.5cm}{\sl\samepage
Physikalisches Institut das RWTH-Aachen, D-52056 Aachen, Germany}
\end{center}\end{sloppypar}
\vspace{2mm}
\begin{sloppypar}
\noindent
R.~Barate,
R.~Bruneli\`ere,
I.~De~Bonis,
D.~Decamp,
C.~Goy,
S.~Jezequel,
J.-P.~Lees,
F.~Martin,
E.~Merle,
\mbox{M.-N.~Minard},
B.~Pietrzyk,
B.~Trocm\'e
\nopagebreak
\begin{center}
\parbox{15.5cm}{\sl\samepage
Laboratoire de Physique des Particules (LAPP), IN$^{2}$P$^{3}$-CNRS,
F-74019 Annecy-le-Vieux Cedex, France}
\end{center}\end{sloppypar}
\vspace{2mm}
\begin{sloppypar}
\noindent
G.~Boix,$^{25}$
S.~Bravo,
M.P.~Casado,
M.~Chmeissani,
J.M.~Crespo,
E.~Fernandez,
M.~Fernandez-Bosman,
Ll.~Garrido,$^{15}$
E.~Graug\'{e}s,
J.~Lopez,
M.~Martinez,
G.~Merino,
R.~Miquel,$^{4}$
Ll.M.~Mir,$^{4}$
A.~Pacheco,
D.~Paneque,
H.~Ruiz
\nopagebreak
\begin{center}
\parbox{15.5cm}{\sl\samepage
Institut de F\'{i}sica d'Altes Energies, Universitat Aut\`{o}noma
de Barcelona, E-08193 Bellaterra (Barcelona), Spain$^{7}$}
\end{center}\end{sloppypar}
\vspace{2mm}
\begin{sloppypar}
\noindent
A.~Colaleo,
D.~Creanza,
N.~De~Filippis,
M.~de~Palma,
G.~Iaselli,
G.~Maggi,
M.~Maggi,
S.~Nuzzo,
A.~Ranieri,
G.~Raso,$^{24}$
F.~Ruggieri,
G.~Selvaggi,
L.~Silvestris,
P.~Tempesta,
A.~Tricomi,$^{3}$
G.~Zito
\nopagebreak
\begin{center}
\parbox{15.5cm}{\sl\samepage
Dipartimento di Fisica, INFN Sezione di Bari, I-70126 Bari, Italy}
\end{center}\end{sloppypar}
\vspace{2mm}
\begin{sloppypar}
\noindent
X.~Huang,
J.~Lin,
Q. Ouyang,
T.~Wang,
Y.~Xie,
R.~Xu,
S.~Xue,
J.~Zhang,
L.~Zhang,
W.~Zhao
\nopagebreak
\begin{center}
\parbox{15.5cm}{\sl\samepage
Institute of High Energy Physics, Academia Sinica, Beijing, The People's
Republic of China$^{8}$}
\end{center}\end{sloppypar}
\vspace{2mm}
\begin{sloppypar}
\noindent
D.~Abbaneo,
P.~Azzurri,
T.~Barklow,$^{30}$
O.~Buchm\"uller,$^{30}$
M.~Cattaneo,
F.~Cerutti,
B.~Clerbaux,$^{34}$
H.~Drevermann,
R.W.~Forty,
M.~Frank,
F.~Gianotti,
T.C.~Greening,$^{26}$
J.B.~Hansen,
J.~Harvey,
D.E.~Hutchcroft,
P.~Janot,
B.~Jost,
M.~Kado,$^{2}$
P.~Mato,
A.~Moutoussi,
F.~Ranjard,
L.~Rolandi,
D.~Schlatter,
G.~Sguazzoni,
W.~Tejessy,
F.~Teubert,
A.~Valassi,
I.~Videau,
J.J.~Ward
\nopagebreak
\begin{center}
\parbox{15.5cm}{\sl\samepage
European Laboratory for Particle Physics (CERN), CH-1211 Geneva 23,
Switzerland}
\end{center}\end{sloppypar}
\vspace{2mm}
\begin{sloppypar}
\noindent
F.~Badaud,
S.~Dessagne,
A.~Falvard,$^{20}$
D.~Fayolle,
P.~Gay,
J.~Jousset,
B.~Michel,
S.~Monteil,
D.~Pallin,
J.M.~Pascolo,
P.~Perret
\nopagebreak
\begin{center}
\parbox{15.5cm}{\sl\samepage
Laboratoire de Physique Corpusculaire, Universit\'e Blaise Pascal,
IN$^{2}$P$^{3}$-CNRS, Clermont-Ferrand, F-63177 Aubi\`{e}re, France}
\end{center}\end{sloppypar}
\vspace{2mm}
\begin{sloppypar}
\noindent
J.D.~Hansen,
J.R.~Hansen,
P.H.~Hansen,
B.S.~Nilsson
\nopagebreak
\begin{center}
\parbox{15.5cm}{\sl\samepage
Niels Bohr Institute, 2100 Copenhagen, DK-Denmark$^{9}$}
\end{center}\end{sloppypar}
\vspace{2mm}
\begin{sloppypar}
\noindent
A.~Kyriakis,
C.~Markou,
E.~Simopoulou,
A.~Vayaki,
K.~Zachariadou
\nopagebreak
\begin{center}
\parbox{15.5cm}{\sl\samepage
Nuclear Research Center Demokritos (NRCD), GR-15310 Attiki, Greece}
\end{center}\end{sloppypar}
\vspace{2mm}
\begin{sloppypar}
\noindent
A.~Blondel,$^{12}$
\mbox{J.-C.~Brient},
F.~Machefert,
A.~Roug\'{e},
M.~Swynghedauw,
R.~Tanaka
\linebreak
H.~Videau
\nopagebreak
\begin{center}
\parbox{15.5cm}{\sl\samepage
Laboratoire de Physique Nucl\'eaire et des Hautes Energies, Ecole
Polytechnique, IN$^{2}$P$^{3}$-CNRS, \mbox{F-91128} Palaiseau Cedex, France}
\end{center}\end{sloppypar}
\vspace{2mm}
\begin{sloppypar}
\noindent
V.~Ciulli,
E.~Focardi,
G.~Parrini
\nopagebreak
\begin{center}
\parbox{15.5cm}{\sl\samepage
Dipartimento di Fisica, Universit\`a di Firenze, INFN Sezione di Firenze,
I-50125 Firenze, Italy}
\end{center}\end{sloppypar}
\vspace{2mm}
\begin{sloppypar}
\noindent
A.~Antonelli,
M.~Antonelli,
G.~Bencivenni,
F.~Bossi,
G.~Capon,
V.~Chiarella,
P.~Laurelli,
G.~Mannocchi,$^{5}$
G.P.~Murtas,
L.~Passalacqua
\nopagebreak
\begin{center}
\parbox{15.5cm}{\sl\samepage
Laboratori Nazionali dell'INFN (LNF-INFN), I-00044 Frascati, Italy}
\end{center}\end{sloppypar}
\vspace{2mm}
\begin{sloppypar}
\noindent
J.~Kennedy,
J.G.~Lynch,
P.~Negus,
V.~O'Shea,
A.S.~Thompson
\nopagebreak
\begin{center}
\parbox{15.5cm}{\sl\samepage
Department of Physics and Astronomy, University of Glasgow, Glasgow G12
8QQ,United Kingdom$^{10}$}
\end{center}\end{sloppypar}
\vspace{2mm}
\begin{sloppypar}
\noindent
S.~Wasserbaech
\nopagebreak
\begin{center}
\parbox{15.5cm}{\sl\samepage
Department of Physics, Haverford College, Haverford, PA 19041-1392, U.S.A.}
\end{center}\end{sloppypar}
\vspace{2mm}
\begin{sloppypar}
\noindent
R.~Cavanaugh,$^{33}$
S.~Dhamotharan,$^{21}$
C.~Geweniger,
P.~Hanke,
V.~Hepp,
E.E.~Kluge,
G.~Leibenguth,
A.~Putzer,
H.~Stenzel,
K.~Tittel,
M.~Wunsch$^{19}$
\nopagebreak
\begin{center}
\parbox{15.5cm}{\sl\samepage
Kirchhoff-Institut f\"ur Physik, Universit\"at Heidelberg, D-69120
Heidelberg, Germany$^{16}$}
\end{center}\end{sloppypar}
\vspace{2mm}
\begin{sloppypar}
\noindent
R.~Beuselinck,
W.~Cameron,
G.~Davies,
P.J.~Dornan,
M.~Girone,$^{1}$
R.D.~Hill,
N.~Marinelli,
J.~Nowell,
S.A.~Rutherford,
J.K.~Sedgbeer,
J.C.~Thompson,$^{14}$
R.~White
\nopagebreak
\begin{center}
\parbox{15.5cm}{\sl\samepage
Department of Physics, Imperial College, London SW7 2BZ,
United Kingdom$^{10}$}
\end{center}\end{sloppypar}
\vspace{2mm}
\begin{sloppypar}
\noindent
V.M.~Ghete,
P.~Girtler,
E.~Kneringer,
D.~Kuhn,
G.~Rudolph
\nopagebreak
\begin{center}
\parbox{15.5cm}{\sl\samepage
Institut f\"ur Experimentalphysik, Universit\"at Innsbruck, A-6020
Innsbruck, Austria$^{18}$}
\end{center}\end{sloppypar}
\vspace{2mm}
\begin{sloppypar}
\noindent
E.~Bouhova-Thacker,
C.K.~Bowdery,
D.P.~Clarke,
G.~Ellis,
A.J.~Finch,
F.~Foster,
G.~Hughes,
R.W.L.~Jones,
M.R.~Pearson,
N.A.~Robertson,
M.~Smizanska
\nopagebreak
\begin{center}
\parbox{15.5cm}{\sl\samepage
Department of Physics, University of Lancaster, Lancaster LA1 4YB,
United Kingdom$^{10}$}
\end{center}\end{sloppypar}
\vspace{2mm}
\begin{sloppypar}
\noindent
O.~van~der~Aa,
C.~Delaere,
V.~Lemaitre
\nopagebreak
\begin{center}
\parbox{15.5cm}{\sl\samepage
Institut de Physique Nucl\'eaire, D\'epartement de Physique, Universit\'e Catholique de Louvain, 1348 Louvain-la-Neuve, Belgium}
\end{center}\end{sloppypar}
\vspace{2mm}
\begin{sloppypar}
\noindent
U.~Blumenschein,
F.~H\"olldorfer,
K.~Jakobs,
F.~Kayser,
K.~Kleinknecht,
A.-S.~M\"uller,
G.~Quast,$^{6}$
B.~Renk,
H.-G.~Sander,
S.~Schmeling,
H.~Wachsmuth,
C.~Zeitnitz,
T.~Ziegler
\nopagebreak
\begin{center}
\parbox{15.5cm}{\sl\samepage
Institut f\"ur Physik, Universit\"at Mainz, D-55099 Mainz, Germany$^{16}$}
\end{center}\end{sloppypar}
\vspace{2mm}
\begin{sloppypar}
\noindent
A.~Bonissent,
P.~Coyle,
C.~Curtil,
A.~Ealet,
D.~Fouchez,
P.~Payre,
A.~Tilquin
\nopagebreak
\begin{center}
\parbox{15.5cm}{\sl\samepage
Centre de Physique des Particules de Marseille, Univ M\'editerran\'ee,
IN$^{2}$P$^{3}$-CNRS, F-13288 Marseille, France}
\end{center}\end{sloppypar}
\vspace{2mm}
\begin{sloppypar}
\noindent
F.~Ragusa
\nopagebreak
\begin{center}
\parbox{15.5cm}{\sl\samepage
Dipartimento di Fisica, Universit\`a di Milano e INFN Sezione di
Milano, I-20133 Milano, Italy.}
\end{center}\end{sloppypar}
\vspace{2mm}
\begin{sloppypar}
\noindent
A.~David,
H.~Dietl,
G.~Ganis,$^{27}$
K.~H\"uttmann,
G.~L\"utjens,
W.~M\"anner,
\mbox{H.-G.~Moser},
R.~Settles,
G.~Wolf
\nopagebreak
\begin{center}
\parbox{15.5cm}{\sl\samepage
Max-Planck-Institut f\"ur Physik, Werner-Heisenberg-Institut,
D-80805 M\"unchen, Germany\footnotemark[16]}
\end{center}\end{sloppypar}
\vspace{2mm}
\begin{sloppypar}
\noindent
J.~Boucrot,
O.~Callot,
M.~Davier,
L.~Duflot,
\mbox{J.-F.~Grivaz},
Ph.~Heusse,
A.~Jacholkowska,$^{32}$
C.~Loomis,
L.~Serin,
\mbox{J.-J.~Veillet},
J.-B.~de~Vivie~de~R\'egie,$^{28}$
C.~Yuan
\nopagebreak
\begin{center}
\parbox{15.5cm}{\sl\samepage
Laboratoire de l'Acc\'el\'erateur Lin\'eaire, Universit\'e de Paris-Sud,
IN$^{2}$P$^{3}$-CNRS, F-91898 Orsay Cedex, France}
\end{center}\end{sloppypar}
\vspace{2mm}
\begin{sloppypar}
\noindent
G.~Bagliesi,
T.~Boccali,
L.~Fo\`a,
A.~Giammanco,
A.~Giassi,
F.~Ligabue,
A.~Messineo,
F.~Palla,
G.~Sanguinetti,
A.~Sciab\`a,
R.~Tenchini,$^{1}$
A.~Venturi,$^{1}$
P.G.~Verdini
\samepage
\begin{center}
\parbox{15.5cm}{\sl\samepage
Dipartimento di Fisica dell'Universit\`a, INFN Sezione di Pisa,
e Scuola Normale Superiore, I-56010 Pisa, Italy}
\end{center}\end{sloppypar}
\vspace{2mm}
\begin{sloppypar}
\noindent
O.~Awunor,
G.A.~Blair,
G.~Cowan,
A.~Garcia-Bellido,
M.G.~Green,
L.T.~Jones,
T.~Medcalf,
A.~Misiejuk,
J.A.~Strong,
P.~Teixeira-Dias
\nopagebreak
\begin{center}
\parbox{15.5cm}{\sl\samepage
Department of Physics, Royal Holloway \& Bedford New College,
University of London, Egham, Surrey TW20 OEX, United Kingdom$^{10}$}
\end{center}\end{sloppypar}
\vspace{2mm}
\begin{sloppypar}
\noindent
R.W.~Clifft,
T.R.~Edgecock,
P.R.~Norton,
I.R.~Tomalin
\nopagebreak
\begin{center}
\parbox{15.5cm}{\sl\samepage
Particle Physics Dept., Rutherford Appleton Laboratory,
Chilton, Didcot, Oxon OX11 OQX, United Kingdom$^{10}$}
\end{center}\end{sloppypar}
\vspace{2mm}
\begin{sloppypar}
\noindent
\mbox{B.~Bloch-Devaux},
D.~Boumediene,
P.~Colas,
B.~Fabbro,
E.~Lan\c{c}on,
\mbox{M.-C.~Lemaire},
E.~Locci,
P.~Perez,
J.~Rander,
B.~Tuchming,
B.~Vallage
\nopagebreak
\begin{center}
\parbox{15.5cm}{\sl\samepage
CEA, DAPNIA/Service de Physique des Particules,
CE-Saclay, F-91191 Gif-sur-Yvette Cedex, France$^{17}$}
\end{center}\end{sloppypar}
\vspace{2mm}
\begin{sloppypar}
\noindent
N.~Konstantinidis,
A.M.~Litke,
G.~Taylor
\nopagebreak
\begin{center}
\parbox{15.5cm}{\sl\samepage
Institute for Particle Physics, University of California at
Santa Cruz, Santa Cruz, CA 95064, USA$^{22}$}
\end{center}\end{sloppypar}
\vspace{2mm}
\begin{sloppypar}
\noindent
C.N.~Booth,
S.~Cartwright,
F.~Combley,$^{31}$
P.N.~Hodgson,
M.~Lehto,
L.F.~Thompson
\nopagebreak
\begin{center}
\parbox{15.5cm}{\sl\samepage
Department of Physics, University of Sheffield, Sheffield S3 7RH,
United Kingdom$^{10}$}
\end{center}\end{sloppypar}
\vspace{2mm}
\begin{sloppypar}
\noindent
K.~Affholderbach,$^{23}$
A.~B\"ohrer,
S.~Brandt,
C.~Grupen,
J.~Hess,
A.~Ngac,
G.~Prange,
U.~Sieler
\nopagebreak
\begin{center}
\parbox{15.5cm}{\sl\samepage
Fachbereich Physik, Universit\"at Siegen, D-57068 Siegen, Germany$^{16}$}
\end{center}\end{sloppypar}
\vspace{2mm}
\begin{sloppypar}
\noindent
C.~Borean,
G.~Giannini
\nopagebreak
\begin{center}
\parbox{15.5cm}{\sl\samepage
Dipartimento di Fisica, Universit\`a di Trieste e INFN Sezione di Trieste,
I-34127 Trieste, Italy}
\end{center}\end{sloppypar}
\vspace{2mm}
\begin{sloppypar}
\noindent
H.~He,
J.~Putz,
J.~Rothberg
\nopagebreak
\begin{center}
\parbox{15.5cm}{\sl\samepage
Experimental Elementary Particle Physics, University of Washington, Seattle,
WA 98195 U.S.A.}
\end{center}\end{sloppypar}
\vspace{2mm}
\begin{sloppypar}
\noindent
S.R.~Armstrong,
K.~Berkelman,
K.~Cranmer,
D.P.S.~Ferguson,
Y.~Gao,$^{29}$
S.~Gonz\'{a}lez,
O.J.~Hayes,
H.~Hu,
S.~Jin,
J.~Kile,
P.A.~McNamara III,
J.~Nielsen,
Y.B.~Pan,
\mbox{J.H.~von~Wimmersperg-Toeller}, 
W.~Wiedenmann,
J.~Wu,
Sau~Lan~Wu,
X.~Wu,
G.~Zobernig
\nopagebreak
\begin{center}
\parbox{15.5cm}{\sl\samepage
Department of Physics, University of Wisconsin, Madison, WI 53706,
USA$^{11}$}
\end{center}\end{sloppypar}
\vspace{2mm}
\begin{sloppypar}
\noindent
G.~Dissertori
\nopagebreak
\begin{center}
\parbox{15.5cm}{\sl\samepage
Institute for Particle Physics, ETH H\"onggerberg, 8093 Z\"urich,
Switzerland.}
\end{center}\end{sloppypar}
}
\footnotetext[1]{Also at CERN, 1211 Geneva 23, Switzerland.}
\footnotetext[2]{Now at Fermilab, PO Box 500, MS 352, Batavia, IL 60510, USA}
\footnotetext[3]{Also at Dipartimento di Fisica di Catania and INFN Sezione di
 Catania, 95129 Catania, Italy.}
\footnotetext[4]{Now at LBNL, Berkeley, CA 94720, U.S.A.}
\footnotetext[5]{Also Istituto di Cosmo-Geofisica del C.N.R., Torino,
Italy.}
\footnotetext[6]{Now at Institut f\"ur Experimentelle Kernphysik, Universit\"at Karlsruhe, 76128 Karlsruhe, Germany.}
\footnotetext[7]{Supported by CICYT, Spain.}
\footnotetext[8]{Supported by the National Science Foundation of China.}
\footnotetext[9]{Supported by the Danish Natural Science Research Council.}
\footnotetext[10]{Supported by the UK Particle Physics and Astronomy Research
Council.}
\footnotetext[11]{Supported by the US Department of Energy, grant
DE-FG0295-ER40896.}
\footnotetext[12]{Now at Departement de Physique Corpusculaire, Universit\'e de
Gen\`eve, 1211 Gen\`eve 4, Switzerland.}
\footnotetext[13]{Supported by the Commission of the European Communities,
contract ERBFMBICT982874.}
\footnotetext[14]{Supported by the Leverhulme Trust.}
\footnotetext[15]{Permanent address: Universitat de Barcelona, 08208 Barcelona,
Spain.}
\footnotetext[16]{Supported by Bundesministerium f\"ur Bildung
und Forschung, Germany.}
\footnotetext[17]{Supported by the Direction des Sciences de la
Mati\`ere, C.E.A.}
\footnotetext[18]{Supported by the Austrian Ministry for Science and Transport.}
\footnotetext[19]{Now at SAP AG, 69185 Walldorf, Germany}
\footnotetext[20]{Now at Groupe d' Astroparticules de Montpellier, Universit\'e de Montpellier II, 34095 Montpellier, France.}
\footnotetext[21]{Now at BNP Paribas, 60325 Frankfurt am Mainz, Germany}
\footnotetext[22]{Supported by the US Department of Energy,
grant DE-FG03-92ER40689.}
\footnotetext[23]{Now at Skyguide, Swissair Navigation Services, Geneva, Switzerland.}
\footnotetext[24]{Also at Dipartimento di Fisica e Tecnologie Relative, Universit\`a di Palermo, Palermo, Italy.}
\footnotetext[25]{Now at McKinsey and Compagny, Avenue Louis Casal 18, 1203 Geneva, Switzerland.}
\footnotetext[26]{Now at Honeywell, Phoenix AZ, U.S.A.}
\footnotetext[27]{Now at INFN Sezione di Roma II, Dipartimento di Fisica, Universit\`a di Roma Tor Vergata, 00133 Roma, Italy.}
\footnotetext[28]{Now at Centre de Physique des Particules de Marseille, Univ M\'editerran\'ee, F-13288 Marseille, France.}
\footnotetext[29]{Also at Department of Physics, Tsinghua University, Beijing, The People's Republic of China.}
\footnotetext[30]{Now at SLAC, Stanford, CA 94309, U.S.A.}
\footnotetext[31]{Deceased.}
\footnotetext[32]{Also at Groupe d' Astroparticules de Montpellier, Universit\'e de Montpellier II, 34095 Montpellier, France.}  
\footnotetext[33]{Now at University of Florida, Department of Physics, Gainesville, Florida 32611-8440, USA}
\footnotetext[34]{Now at Institut Inter-universitaire des hautes Energies (IIHE), CP 230, Universit\'{e} Libre de Bruxelles, 1050 Bruxelles, Belgique}
\setlength{\parskip}{\saveparskip}
\setlength{\textheight}{\savetextheight}
\setlength{\topmargin}{\savetopmargin}
\setlength{\textwidth}{\savetextwidth}
\setlength{\oddsidemargin}{\saveoddsidemargin}
\setlength{\topsep}{\savetopsep}
\normalsize
\newpage
\pagestyle{plain}
\setcounter{page}{1}

\pagebreak
\pagestyle{plain}
\setcounter{page}{1}
\normalsize
\section{Introduction}

The results of searches for scalar quarks with the data collected 
in the year 2000 by the ALEPH detector at LEP are presented in this
letter. The energies and integrated luminosities of the analysed data
samples are given in Table~\ref{tab:lumi}. 
Previous results obtained with lower energy data have been reported by
ALEPH in Refs.~[1--5] and by the other LEP collaborations in Refs.~[6--8].
%
%
\begin{table}[htbp]
\begin{center}
\caption{\small Integrated luminosities, centre-of-mass energy
   ranges and mean centre-of-mass energy
   values for the data collected by the ALEPH detector in the year 2000. }
\begin{tabular}{|c|c|c|} 
  \multicolumn{3}{c}{} \\  
\hline 
\rule{0pt}{4mm}
 Luminosity $[\ipb]$ & Energy range [GeV] & $\langle \sqrt{s} \rangle$ [GeV] \\ \hline \hline
     9.4            & $207-209$        & 208.0 \\
   122.6            & $206-207$        & 206.6 \\
    75.3            & $204-206$        & 205.2 \\ \hline
\end{tabular}
\label{tab:lumi}
\end{center}
\end{table}

The theoretical framework for these studies is the supersymmetric
extension of the Standard Model~\cite{mssm}, with R-parity
conservation. The lightest supersymmetric particle (LSP) is assumed to
be the lightest neutralino $\chi$ or the sneutrino $\snu$. Such an LSP
is stable and weakly interacting. Each chirality state of the Standard
Model fermions has a scalar supersymmetric partner. The scalar quarks
(squarks) $\sqL$ and $\sqR$ are the supersymmetric partners of the
left-handed and right-handed quarks, respectively. The mass
eigenstates are orthogonal combinations of the weak interaction
eigenstates $\sqL$ and $\sqR$. The mixing angle $\thsqua$ is defined
in such a way that $\squa = \sqL \cos \thsqua + \sqR \sin \thsqua$ is
the lighter squark. The off-diagonal terms of the mass matrix,
responsible for mixing, read, with standard notation: $m_{\mathrm q}
(A_{\mathrm q} - \mu\kappa$), with $\kappa = \tan\beta$ for down-type
and $\kappa = 1/\tan\beta$ for up-type quarks. Since the size of this
mixing term is proportional to the mass of the Standard Model partner,
it could well be that the lightest supersymmetric charged particle is
the lighter scalar top (stop, $\sto$) or, in particular for large
$\tan\beta$ values, the lighter scalar bottom (sbottom,
$\sbot$). Squarks could be produced at LEP in pairs, $\ee  \to \squa
\bar{\squa}$, via $s$-channel exchange of a virtual photon or Z. The
production cross section~\cite{DrHiKo} depends on $\thsqua$ when
mixing is relevant, {\em i.e.}, for stops and sbottoms.

\begin{figure}[t]
  \begin{center}
        \begin{picture}(0,0)
        \put(18,15){\mbox{\bf (a)}}
        \end{picture}
    \epsfig{file=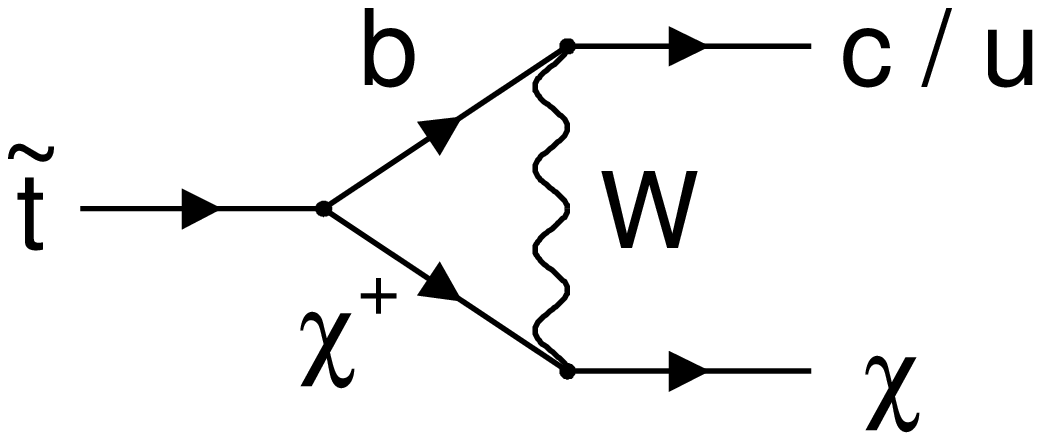,width=0.3\textwidth}
    \hskip 0.01\textwidth
        \begin{picture}(0,0)
        \put(18,15){\mbox{\bf (b)}}
        \end{picture}
    \epsfig{file=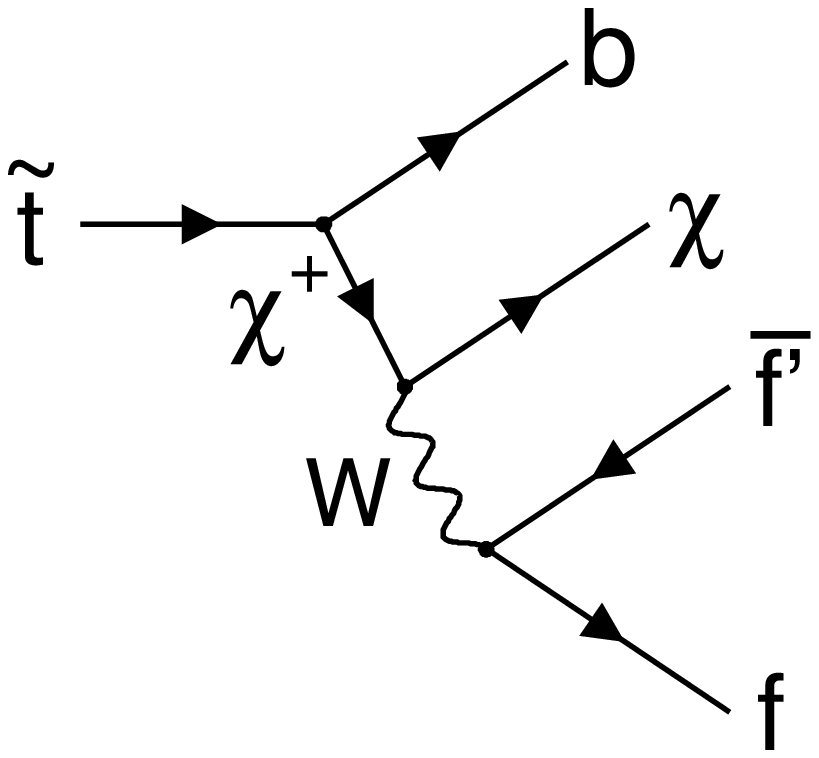,width=0.237\textwidth}
    \hskip 0.01\textwidth
        \begin{picture}(0,0)
        \put(18,15){\mbox{\bf (c)}}
        \end{picture}
    \epsfig{file=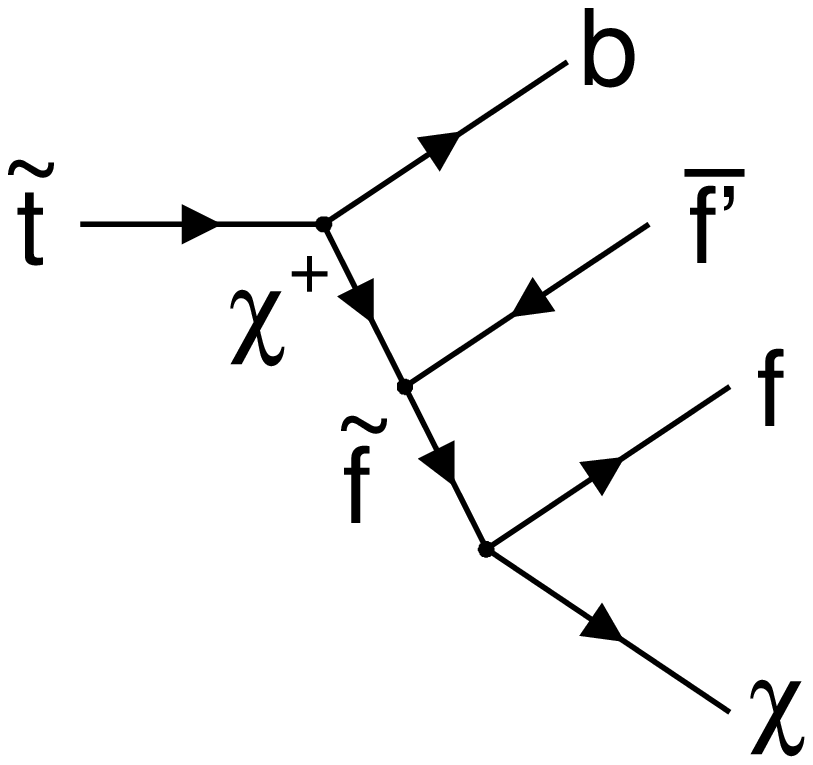,width=0.237\textwidth}\\
    \hskip 0.03\textwidth
        \begin{picture}(0,0)
        \put(18,0){\mbox{\bf (d)}}
        \end{picture}
    \epsfig{file=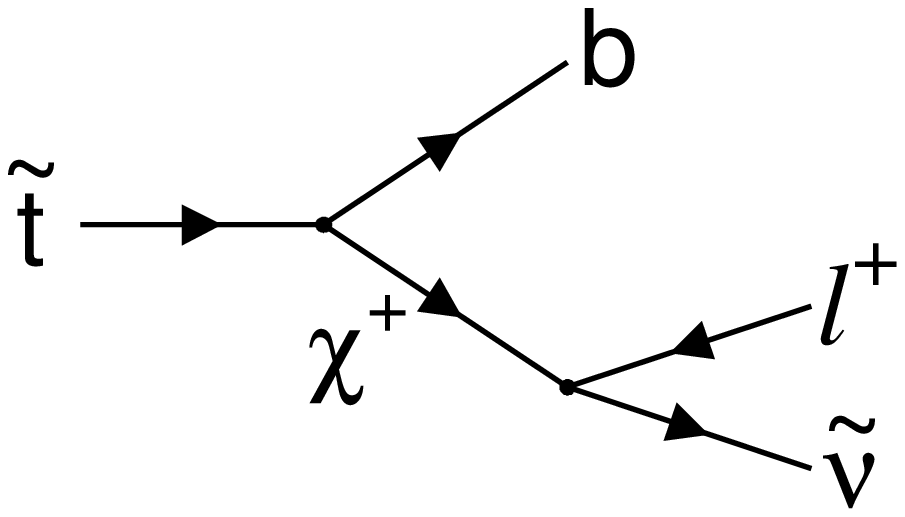,width=0.258\textwidth}
    \hskip 0.08\textwidth
        \begin{picture}(0,0)
        \put(18,0){\mbox{\bf (e)}}
        \end{picture}
    \epsfig{file=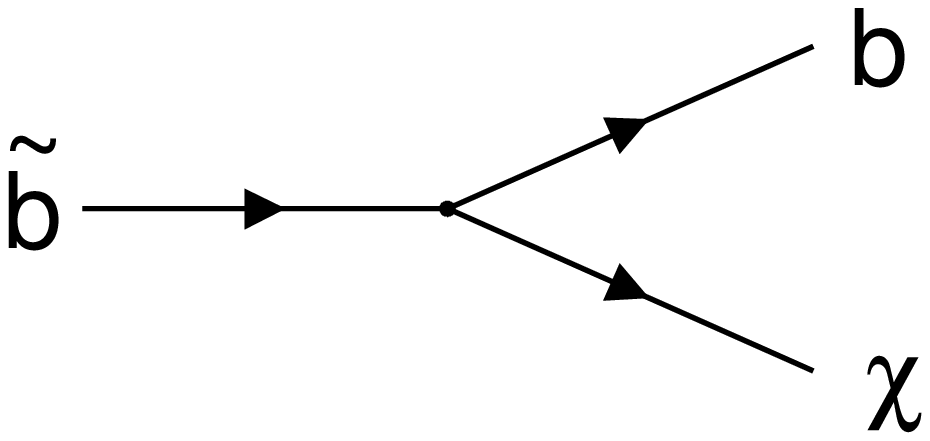,width=0.27\textwidth}
    \caption{\small Squark decay diagrams considered in this letter:
    (a) $\sto \to \charm/\u \, \chi$; (b) $\fbody$ via W exchange and 
    (c) $\fbody$ via sfermion exchange; (d) $\sto \to \b \ell \snu$; 
    (e) $\sbot\to \b \chi$.}
    \label{fig:feyn}
  \end{center}
\end{figure}

The searches for stops described here assume that all supersymmetric
particles except the lightest neutralino $\chi$ and possibly the sneutrino
$\snu$ are heavier than the stop. 
Under these assumptions, the allowed decay channels
are $\sto \to \mathrm{c} / \mathrm{u} \, \chi$,  $\fbody$ and $\sto
\to \b \ell \snu$~\cite{DrHiKo,BoDjMa}. The corresponding diagrams are
shown in Fig.~\ref{fig:feyn}.
The decay $\sto \to \mathrm{c} \chi$ (Fig.~\ref{fig:feyn}a) 
proceeds only via loops and has a very small width, of the order of
0.01--$1\eV$~\cite{DrHiKo}, depending on the mass difference $\dM$
between the stop and the neutralino, and on the masses and field
content of the particles involved in the loops. For low enough $\dM$
values ($\Delta M \lesssim 6\GeVcc$), the stop lifetime
becomes sizeable, and must be taken into account in the searches for
stop production. If $\dM$ is so small that the $\sto \to \charm \chi$
channel is kinematically closed, the dominant decay mode becomes
$\sto \to \u \chi$, and the stop can then be considered as stable for
practical purposes.

\pagebreak

For stop masses accessible at LEP, {\it i.e.}, $\lesssim 100 \GeVcc$,
the decay mode $\fbody$ is mediated by virtual chargino and W
(Fig.~\ref{fig:feyn}b) or sfermion (Fig.~\ref{fig:feyn}c) exchange. It
is therefore of the same order in perturbation theory as the
loop-induced $\sto \to \charm \chi$ decay, and can be substantially
enhanced if charginos have masses not much larger than their present
experimental bounds, and could even dominate for light
sfermions~\cite{BoDjMa}. The four-body decay channel yields topologies
with b-jets, additional jets and/or leptons, and with missing mass and
missing energy. A new multi-jet analysis, hereafter called MJ, has
been designed to cope with these final states.

The $\sto \to \b \ell \snu$ channel proceeds via virtual chargino
exchange (Fig.~\ref{fig:feyn}d) and has a width of the order of
0.1--$10\keV$~\cite{DrHiKo}. This decay channel dominates when it is
kinematically allowed, {\it i.e.}, if the lightest $\snu$ is lighter than the
stop. If the lightest neutralino is the LSP, the sneutrino decays invisibly
into $\chi \nu$ without any change in the experimental topology.

Under the assumption that the $\sbot$ is lighter than all supersymmetric
particles except the $\chi$, the $\sbot$ will decay as $\sbot \to \b
\chi$ (Fig.~\ref{fig:feyn}e). Compared to the $\sto$, the
$\sbot$ decay width is larger, of the order of 10--$100\MeV$. 

The supersymmetric partners of the light quarks are generally expected
to be quite heavy. If they are light enough to be within the reach of
LEP, their dominant decay mode is expected to be $\squa \to \q \neu$.

The final state topologies addressed by the searches presented in this
letter are summarised in Table~\ref{tab:topo}, together with the related
signal processes and with the references where analysis details can be
found.

\begin{table}[htbp]
  \caption{\small 
    Topologies studied in the different scenarios.
    \label{tab:topo}}
  \begin{center}
    \begin{tabular}{|c|l|l|l|}
      \hline
      Production & Decay mode & Topology/Analysis & References \\
      \hline
      \hline
      \rule{-4pt}{14pt} 
      $\sto\bar{\sto}$   & $\sto \to \charm \chi$ ($\Delta M \gtrsim 6
      \GeVcc$) & Acoplanar
      jets (AJ) &  \cite{ALEPHsqone,ALEPHsqtwo,ALEPHsqthree,ALEPHsqfour} \\
      $\sto\bar{\sto}$   & $\sto \to \charm/ \u \, \chi$ ($\Delta M \lesssim 6
      \GeVcc$) & Long-lived hadrons &  \cite{llh} \\
      $\sto\bar{\sto}$   & $\fbody$   &
      Multi-jets (MJ) & This letter\\
      $\sto \bar{\sto}$  & $\sto \to \b \ell \snu$ & AJ
      plus leptons &  \cite{ALEPHsqone,ALEPHsqtwo,ALEPHsqthree,ALEPHsqfour} \\      
      $\sbot\bar{\sbot}$ & $\sbot \to \b \chi$      & AJ
      plus b tagging &  \cite{ALEPHsqone,ALEPHsqtwo,ALEPHsqthree,ALEPHsqfour} \\
      $\squa\bar{\squa}$ & $\squa \to \q \chi$      & AJ &
      \cite{ALEPHsqtwo,ALEPHsqthree,ALEPHsqfour} \\
      \hline
    \end{tabular}
  \end{center}
\end{table}

This letter is organised as follows. In Section~2, the ALEPH detector and 
the simulated samples
used for the analyses are described. Section~3 is dedicated to the 
selection algorithms with emphasis on the new search 
for four-body stop decays.
In Section~4 the results of the searches 
are given, along with their interpretation in the theoretical framework. 
The conclusions of the letter are given in Section~5. 

\section{ALEPH detector and event simulation}

A thorough description of the ALEPH detector and of its performance, 
as well as of the standard reconstruction and analysis algorithms,
can be found in Refs.~\cite{Alnim,Alperf}. Only a brief summary is
given here.  

The trajectories of charged particles are measured by a silicon 
vertex detector (VDET), a cylindrical multi-wire drift chamber (ITC)
and a large time projection chamber (TPC).
These detectors are immersed in an axial magnetic field of $1.5 \T$ 
provided by a superconducting solenoidal coil. The VDET consists of
two cylindrical layers of silicon microstrip detectors; it performs
precise measurements of the impact parameter in space, yielding
powerful short-lived particle tags, as described in Ref.~\cite{Btag}.

The electromagnetic calorimeter (ECAL), placed between the TPC and the
coil, is a highly-segmented sandwich of lead planes and proportional
wire chambers. It consists of a barrel and two endcaps. The hadron
calorimeter (HCAL) consists of the iron return yoke of the magnet
instrumented with streamer tubes. It is surrounded by two double
layers of streamer tubes, the muon chambers. The luminosity monitors
(LCAL and SiCAL) extend the calorimeter hermeticity down to 34 mrad
from the beam axis.

The energy flow algorithm described in Ref.~\cite{Alperf} combines the 
measurements of the tracking detectors and of the calorimeters 
into  ``objects'' classified as charged particles, photons, and
neutral hadrons. The energy resolution achieved with this algorithm is 
$(0.6\sqrt E + 0.6)$~GeV ($E$ in GeV).
Electrons are identified by comparing the energy deposit in ECAL to 
the momentum measured in the tracking system, by using the shower profile 
in the electromagnetic calorimeter, and by the measurement of the specific 
ionization energy loss in the TPC. 
The identification of muons makes use of the hit pattern
in HCAL and of the muon chambers.

Signal event samples were simulated with the generator described in
Ref.~\cite{ALEPHsqone} for $\sto \to \charm \chi$, $\sbot \to \b \chi$,
$\squa \to \q \chi$ and $\sto \to \b \ell \snu$. A modified version of
this generator was designed to simulate the channel $\fbody$, where
the final state is modelled according to phase space and including
parton shower development. The generation of $\sto \to \charm / \u \,
\chi$ with lifetime follows the procedure described in Ref.~\cite{llh}.

To simulate the relevant Standard Model background processes, several 
Monte Carlo generators were used: BHWIDE~\cite{bhwide} for Bhabha
scattering, KORALZ~\cite{koralz} for $\mu^+\mu^-$ and $\tau^+\tau^-$
production, PHOT02~\cite{phot02} for $\gaga$ interactions,
KORALW~\cite{koralw} for WW production, and PYTHIA~\cite{jetset} for
the other processes ($\ee\to\q\bar{\q}(\gamma)$, We$\nu$, Zee, ZZ,
Z$\nu\bar\nu$). The sizes of the simulated samples typically
correspond to ten times the integrated luminosity of the data.

All background and signal samples were processed through the full detector 
simulation.

\section{Event selections}

Several selection algorithms have been developed to search for the
topologies given in Table~\ref{tab:topo}. All these channels are
characterised by missing energy. The event properties depend
significantly on $\dM$, the mass difference between the decaying
squark and the $\chi$ (or the $\snu$ in the case of $\sto \to
\b\ell\snu$). When $\dM$ is large, there is a substantial amount of
visible energy, and the signal events tend to look like WW, We$\nu$,
ZZ, and $\q\bar{\q}(\gamma)$ events. When $\dM$ is small, the visible
energy is small, and the signal events are therefore similar to
$\gamma\gamma$ events. In order to cope with the different signal
topologies and background situations, each analysis employs selections
dependent on the $\dM$ range. The stop lifetime may become sizeable at
small $\dM$, in which case the signal final state topology depends
strongly on the $\sto$ decay length $\lambda_\sto$; three different
selections are used, each designed to cope with a specific
$\lambda_\sto$ range~\cite{llh}.

The optimisation of the selection criteria as well as the best
combination of selections as a function of $\dM$ and
$\lambda_\sto$ were obtained according to the $\nnc$
prescription~\cite{nbar95}, {\it i.e.}, by minimisation of the
95\%~C.L. cross section upper limit expected in the absence of a
signal. The selections are mostly independent of the centre-of-mass
energy except for an appropriate rescaling of the cuts with $\sqrt{s}$
when relevant. The selections applied to the year 2000 data follow
closely those described in Refs.~[1--5]
except for the new analysis developed to address the $\fbody$ decay,
hereafter described in some detail.

\mathversion{bold}
\subsection{Search for $\fbody$}
\mathversion{normal}

The MJ analysis consists of a small, a large and a very large $\dM$
selection. These selections are designed to address simultaneously all
$\b \neu \q \bar{\q}'$ and $\b \neu \ell \nu$ final states,
independently of the decay branching ratios. The selections use
several anti-$\gaga$ criteria, reported in Table~\ref{tab:4bgg}. The
cuts are derived from the AJ selection, described in 
Ref.~\cite{ALEPHsqone} as well as the variables used. Only the
relevant differences are discussed in the following.

\renewcommand{\arraystretch}{1.1}
\begin{table}[htbp]
  \caption{\small
    Criteria used in the MJ selections to address the backgrounds from
    (A) $\ggqq$, (B) $\ggqq$ with spurious calorimetric objects and
    (C) $\ggtt$. The $^\dagger$ indicates that the cut is applied when
    the azimuthal angle of the missing momentum $\phimiss$ is within
    $15\gradi$ of the vertical plane. \label{tab:4bgg}}
  \begin{center}
    \begin{tabular}{|c||c||p{5cm}|p{5cm}|}
\hline 
  &      & Small $\dM$ & Large and Very large $\dM$ \\
\hline \hline
A & $\nch$           & \multicolumn{2}{|c|}{$> 10$} \\ \cline{2-4}
  & $\mvis$          & \multicolumn{2}{|c|}{$> 10 \GeVcc$} \\ \cline{2-4}  
  & $\etwelve$       & $=0$                   & $< 0.05 \sqrt{s}$ \\ \cline{2-4} 
  & $\ethirty$       & $< 0.25 \etot$         & $< 0.3 \etot$   \\ \cline{2-4}
  & $\acop$ {\small (acoplanarity)} & $<172.5\gradi$         & $<174\gradi$     \\\cline{2-4}
  & $\acopt$ {\small (transverse acop.)}         & $<175\gradi$           & $<175\gradi$     \\\cline{2-4}
  & $\pt/\sqrt{s}$   & $>0.005$ ($>0.01$)$^\dagger$      & $>0.05$ ($>0.075$)$^\dagger$ \\\cline{2-4}
  & $\pt/\etot$      & $>1.305-0.00725\acop$  & $>0.2$    \\ \cline{2-4}
  & $\mmis$          & $< 25.0\etot$          & \\ \cline{2-4}
  & $\thpoint$       & $> 15\gradi$           & $> 5\gradi$ if $\thscat < 15\gradi$ \\\cline{2-4}
  & $|\cos \thmiss|$ & $<0.8$                 & $<0.95$                       \\ \cline{2-4}
  & $|\cos \ththr|$  & $<0.75$                &                              \\ \cline{2-4}
  & $\mhad$          & $ > 10 \GeVcc$         & \\
 \hline          
 \hline          
B & $\ptnoNH$        & $> 0.01\sqrt{s}$         & $< 0.03 \sqrt{s}$ if $\ehad>0.45\evis$ \\ \cline{2-4}
  & $\ptch$          & $> 0.005 \sqrt{s}$       &                                       \\ \cline{2-4}
  & ${E^{\rm NH}_{\rm 1}}$ &  $< 0.3\sqrt{s}$ & \\ \cline{2-4}
  & $\ehad$          &                          & $<0.3\evis$ \\ \cline{2-4}
  & $\ewedge$        &                          & $<0.075\sqrt{s}$ \\ 
\hline
\hline
C & $\nch^{{\rm jet}_i}$, $i=1,2$   & $> 4$ & \\ \cline{2-4}
  & $m^{{\rm jet}_i}$, $i=1,2$      & $> 4\GeVcc$ & \\ \hline
    \end{tabular}
  \end{center}
\end{table}
\renewcommand{\arraystretch}{1.}

In the $\fbody$ channel, the b quark in the final state produces a
visible mass higher than in the $\sto \to \charm \chi$
channel. Therefore, for the small $\dM$ selection, the cut on the
number of charged particle tracks $\nch$ is reinforced by requiring
$\nch>10$, and both the visible mass, $\mvis$,
and the visible mass computed excluding the leading lepton,
$\mhad$, are required to be greater than $10 \GeVcc$. 
These tighter cuts allow others to be loosened: the transverse
momentum $\pt$ and that calculated excluding the neutral hadrons,
$\ptnoNH$, must be greater than $0.005\sqrt{s}$ and $0.01\sqrt{s}$,
respectively. The remaining background is reduced in the small $\dM$
selection by requiring the thrust to be smaller than 0.875, and by the
cut $\etot<0.26\sqrt{s}$.

For the large $\dM$ selections, the multi-jet signature is addressed
by requiring $y_{45}$, as calculated with the DURHAM
algorithm~\cite{durham}, to be greater than 0.001. The level of the
WW, ZZ and We$\nu$ background is reduced by taking advantage of the 
b-quark content in the $\fbody$ final state.
The value of $-\log_{10} P_{\rm uds}$ is required to be greater than
0.5, where $P_{\rm uds}$ is the b-tag event probability introduced in
Ref.~\cite{Btag}. This background is further suppressed by a missing
mass cut, the location of which is a function of the $\dM$ of the
signal considered. For example, for $\dM=20$, 30 and $40\GeVcc$ the 
optimal cuts are $\mmis/\sqrt{s}>0.75$, 0.70 and 0.65, respectively.

The region where the very large $\dM$ selection applies is
characterised by a higher visible mass. The sliding cut on the missing
mass is looser than that in the large $\dM$ selection. For $\dM=20$,
40 and $60\GeVcc$ the optimal cuts are $\mmis/\sqrt{s}>0.58$, 0.34 and
0.10. Other cuts are then necessary to reduce the background mainly
due to WW events. Similarly to the large $\dM$ case, $-\log_{10}
P_{\rm uds}$ and $y_{45}$ are required to be greater than 0.5 and
0.003, respectively. The mean momentum of all reconstructed charged
particle tracks must be less than $0.007\sqrt{s}$. To reduce
background from semileptonic W decays, the fraction of visible energy
due to charged objects excluding the leading lepton is required to be
greater than $0.5$, and the leading lepton, if present, must not be
isolated, {\em i.e.}, the additional energy deposited in a $30^\circ$
cone around its direction must be at least $50\%$ of its energy. At
this level, the remaining background consists of WW events with energy
lost in the beam pipe, responsible for the missing mass. These events
are rejected by requiring $|p_{z}|<0.1\sqrt{s}$ and the energy
$\etwelve$ deposited at polar angle smaller than $12^\circ$ to be 
less than $0.015 \sqrt{s}$.
 
The efficiencies of the three selections were parametrised as a
function of $\dM$ for each stop pair final state that may result
from the decay channels considered ($\sto \to \charm \neu$, $\sto
\to \b \neu \ell \nu$, $\sto \to \b \neu \q \bar{\q}'$). 
This allows the signal efficiency to be parametrised as a function of
the branching ratios. The efficiencies were checked to be practically
independent of the lepton flavour ($\e$, $\mu$, $\tau$) in the $\sto
\to \b \neu \ell \nu$ decay. The small, large and very
large $\dM$ selections are combined using the $\nnc$ procedure as a
function of $\dM$ and of the branching ratios. The background to the
small $\dM$ selection is dominated by $\ggqq$ events and has a total
expectation of $5.0\fb$, while the backgrounds of the large and very
large $\dM$ selections, dominated by WW and other four-fermion
processes amount to 3.5 and $4.4\fb$, respectively.  

\subsection{Systematic uncertainties}

The efficiencies of the MJ analysis may be affected by uncertainties
regarding the assumptions on the stop hadron physics and by
uncertainties related to the detector response. The results of the
systematic studies are summarised in Table~\ref{stsys} for the three
selections.

 \begin{table}[htbp]
 \caption{\small Summary of the relative systematic uncertainties on
  the efficiencies of the MJ analysis.}
 \label{stsys}
 \begin{center}
 \begin{tabular}{|l|c|c|c|}           \hline
   \multicolumn{4}{|c|}{ Systematic uncertainties (\%)} \\ \hline\hline
   \multicolumn{1}{|c|}{} & \multicolumn{3}{c|}{MJ selections} \\ 
                                               \cline{2-4}
 
         & Small $\dM$ & Large $\dM$ & Very large $\dM$\\
                                               \cline{2-4}

  $\mx$       (0.3--$1.0\GeV$)    & 9 & 2 & 3 \\

  $\epsilon_{\sto}$ ($10^{-5}$--$10^{-4}$)
                                   & 2 & 2 & 2 \\
  $\thstop$ (0$^{\circ}$--56$^{\circ}$) 
                                   & 3 & 1 & 1 \\

  Detector and reconstruction      & 2 & 1 & 2 \\

  Monte Carlo statistics           & 3  & 3 & 3 \\ \hline

   TOTAL                           & 10 & 4 & 5 \\ \hline
\end{tabular}
\end{center}
\end{table}

The systematic effects from the assumptions on the stop 
hadron physics were assessed by varying the parameters of the model 
implemented in the generator as in Ref.~\cite{ALEPHsqone}.
The uncertainties from the stop hadron mass were evaluated by varying 
the effective spectator mass $\mx$, set to $0.5\GeVcc$ 
in the analysis, in the range between 0.3 and $1.0\GeVcc$. 
The efficiencies of the large and very large $\dM$ selections are
almost insensitive to this change. The 9$\%$ effect found for the
small $\dM$ selection reflects the variation in the invariant mass
available for the hadronic system.

The systematic error due to the uncertainty on the stop fragmentation
was evaluated by varying $\epsilon_\sto$ by an order of magnitude,
where $\epsilon_\sto$ is the parameter of the Peterson fragmentation
function~\cite{peterson}. The effect on the efficiency is very small
($\sim 2\%$).

The amount of initial state radiation in stop pair production depends on the 
value of the stop coupling to the Z boson, which is controlled by the stop 
mixing angle. A variation of $\thstop$ from $56^{\circ}$ to
$0^{\circ}$, {\em i.e.}, from minimal to maximal coupling, was
applied. The effect was found to be small in all selections, at the
level of 1 to 3\%.

Detector effects have been studied for the variables used in the 
selections. The distributions of all relevant variables show good
agreement with the simulation. In particular, the b-tagging performance
was checked on hadronic events collected at the Z resonance. The
systematic errors associated to detector effects and to the
reconstruction procedure were found to be negligible.

Beam-related background, not included in the event simulation, may affect 
the $\etwelve$ variable. Its effect on the selection efficiency was
determined from data collected at random beam crossings. The net
effect is a relative decrease of the signal efficiency by about
5\%. The uncertainty on this correction is negligible.

Finally, an additional uncertainty of $3\%$ due to the limited Monte Carlo 
statistics was added. The total systematic uncertainty is at the level
of 10\% for the small $\dM$ selection. It is dominated by the limited
knowledge of the stop hadron physics, and results from rather extreme
changes in the model parameters. The systematic uncertainties for the
large and very large $\dM$ selections are at the level of 4--5$\%$.  

The systematic uncertainties in the selections other than for the 
$\fbody$ channel are essentially identical to those reported in 
Refs.~\cite{llh,ALEPHsqfour}.

\section{Results and interpretation}

The numbers of candidate events selected and background events
expected are reported in Table~\ref{tab:res} for all the data samples
used to derive the results below. An overall agreement is observed. In
particular, a total of six candidate events is selected by the new
MJ analysis, with 8.5 events expected from background processes; two
events are found by each of the selections, in agreement with
predictions of 3.3, 2.3 and 2.9 background events at small, large and 
very large $\dM$, respectively.

\pagebreak

\renewcommand{\arraystretch}{1.2}
\begin{table}[t]
  \caption{\small 
    Numbers of candidate events observed ($N_{\rm obs}$) 
    and expected from background ($N_{\rm exp}$) for the different
    selections. Also given are the sizes ($\int \cal{L}$dt) and the
    average centre-of-mass energies $\langle\sqrt{s}\rangle$ of the
    samples analysed.}
  \label{tab:res}
  \begin{center}
    \begin{tabular}{|c|c||c|c|c|c|c|c|c|c|}
      \hline
      & Year    & 
      \multicolumn{2}{c|}{1997} &
      \multicolumn{2}{c|}{1998} & 
      \multicolumn{2}{c|}{1999} &
      \multicolumn{2}{c|}{2000} \\ 
Sample & $\int {\cal L} {\rm dt}$ $[\ipb]$ & 
      \multicolumn{2}{c|}{57.0} &
      \multicolumn{2}{c|}{173.6} & 
      \multicolumn{2}{c|}{236.9} &
      \multicolumn{2}{c|}{207.3} \\ 
      &  $\langle \sqrt{s} \rangle$ $[\GeV]$ &
      \multicolumn{2}{c|}{182.7} &
      \multicolumn{2}{c|}{188.6} & 
      \multicolumn{2}{c|}{197.6} &
      \multicolumn{2}{c|}{206.2} \\ 
      \hline
      \hline
      Analysis & Selection & 
      $N_{\rm obs}$ & $N_{\rm exp}$ &
      $N_{\rm obs}$ & $N_{\rm exp}$ &
      $N_{\rm obs}$ & $N_{\rm exp}$ &
      $N_{\rm obs}$ & $N_{\rm exp}$ \\
      \hline
      \hline
      AJ & small $\dM$           & 1 & 1.5 & 3 & 5.5 & 2 & 2.4 & 2  & 2.1 \\
                     & large $\dM$            & 4 & 3.5 & 5 & 4.0 & 8 & 7.3 & 11 & 8.6 \\
      \hline                                                
      Long-lived     & small $\lambda_{\sto}${\white .}$\!\!$ & \multicolumn{8}{c|}{AJ, small $\dM$}\\
      \cline{3-10}
      hadrons        & intermediate $\lambda_{\sto}${\white .}$\!\!$ & - & - & 0 & 0.3 & 0 & 0.5 & 0 & 0.4 \\
                     & large $\lambda_\sto${\white .}$\!\!$  &  -  & - & 1 & 0.4 & 0 & 0.6 & 0 & 0.6 \\
      \hline                                                
      MJ     & small $\dM$           & 0 & 0.3 & 1 & 0.7 & 1 & 1.2 & 0 & 1.1 \\
                     & large $\dM$           & 1 & 0.2 & 0 & 0.6 & 0 & 0.8 & 1 & 0.7 \\
                     & very large $\dM$      & 0 & 0.2 & 0 & 0.8 & 1 & 1.0 & 1 & 0.9 \\
      \hline                                                
      AJ & small $\dM$           & 1 & 0.8 & 0 & 1.9 & 3 & 2.6 & 0 & 2.4 \\
      plus leptons   & large $\dM$           & 0 & 0.1 & 2 & 0.4 & 2 & 1.4 & 3 & 1.6 \\
      \hline
      AJ & small $\dM$           & 0 & 1.1 & 3 & 3.3 & 1 & 2.2 & 2 & 2.0 \\
      plus b tagging & large $\dM$           & 1 & 0.6 & 0 & 0.9 & 1 & 0.7 & 0 & 1.2 \\
      \hline                                                
    \end{tabular}
  \end{center}
\end{table}
\renewcommand{\arraystretch}{1.}

In the framework of the supersymmetric extension of the Standard
Model~\cite{mssm}, the outcome of these searches can be translated
into constraints in the space of the relevant parameters. In this
process the systematic uncertainties on the 
selection efficiencies were included according to the method described
in Ref.~\cite{syse}, and no background subtraction was applied. The
constraints discussed below, derived from the results given in
Table~\ref{tab:res}, are at 95\% confidence level.

\begin{figure}[p]
  \begin{center}
        \begin{picture}(0,0)
        \put(18,20){\mbox{\bf (a)}}
        \end{picture}
    \epsfig{file=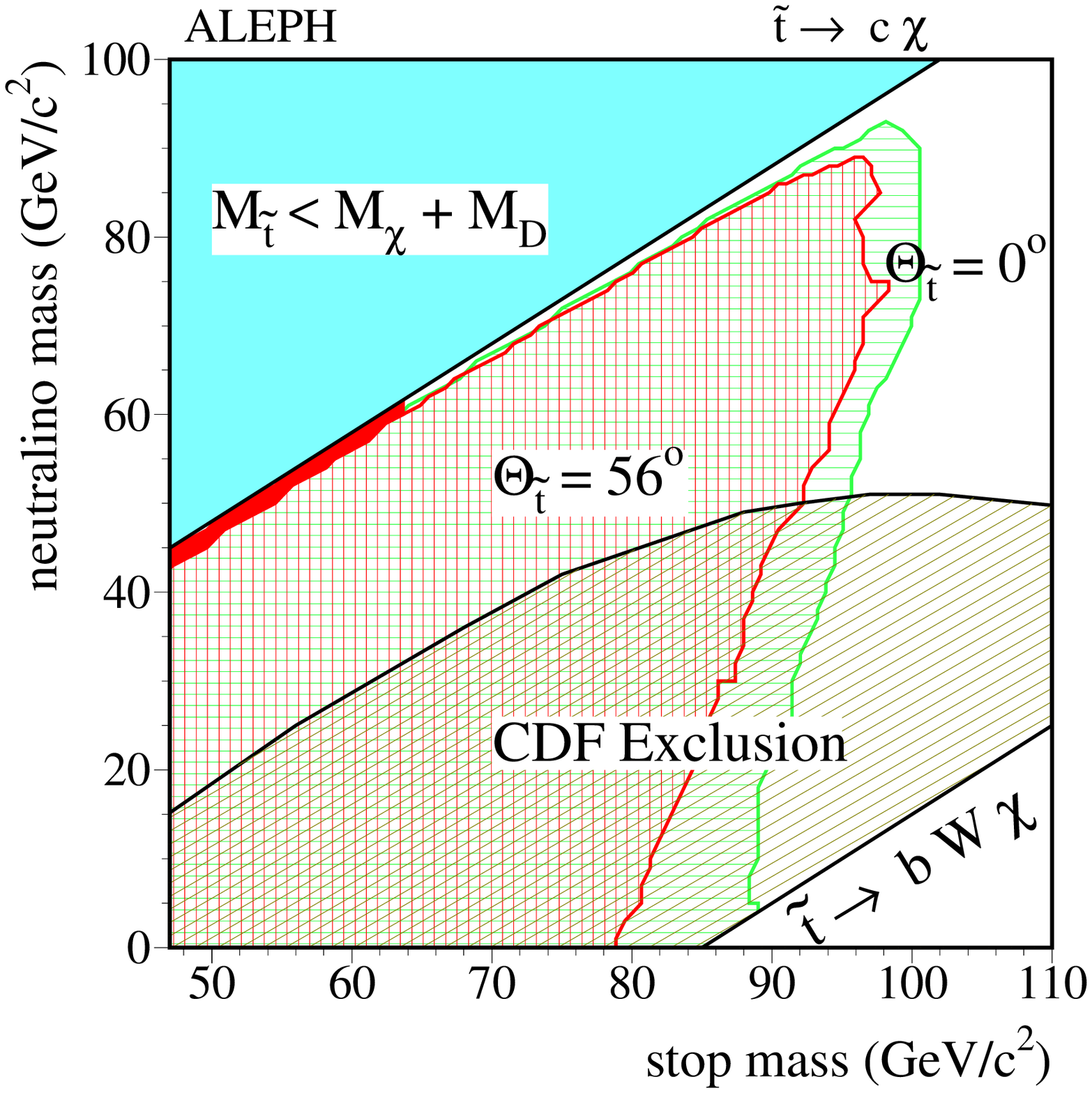,width=0.55\textwidth} \\
        \begin{picture}(0,0)
        \put(18,20){\mbox{\bf (b)}}
        \end{picture}
    \epsfig{file=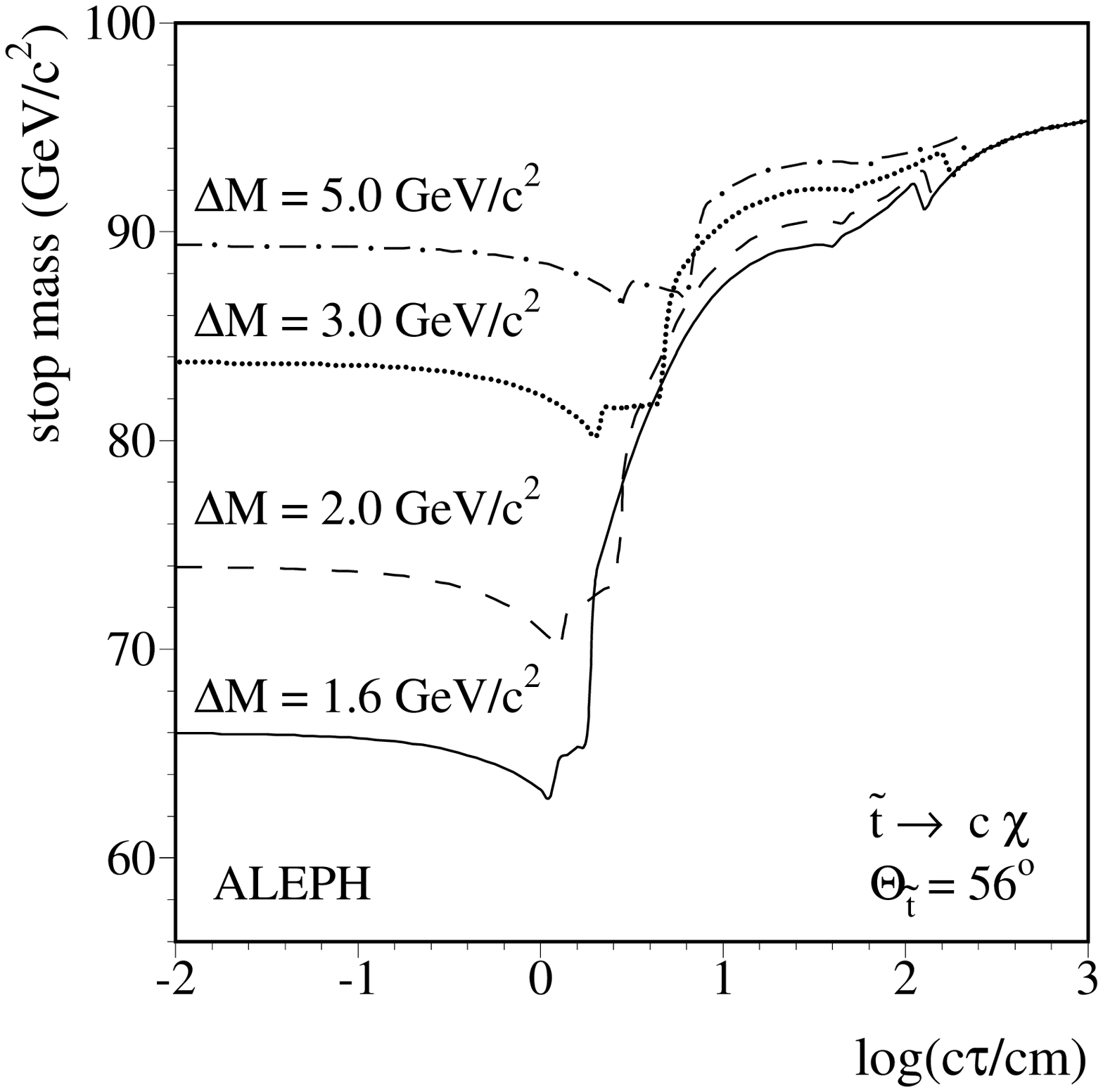,width=0.485\textwidth}
        \begin{picture}(0,0)
        \put(18,20){\mbox{\bf (c)}}
        \end{picture}
    \epsfig{file=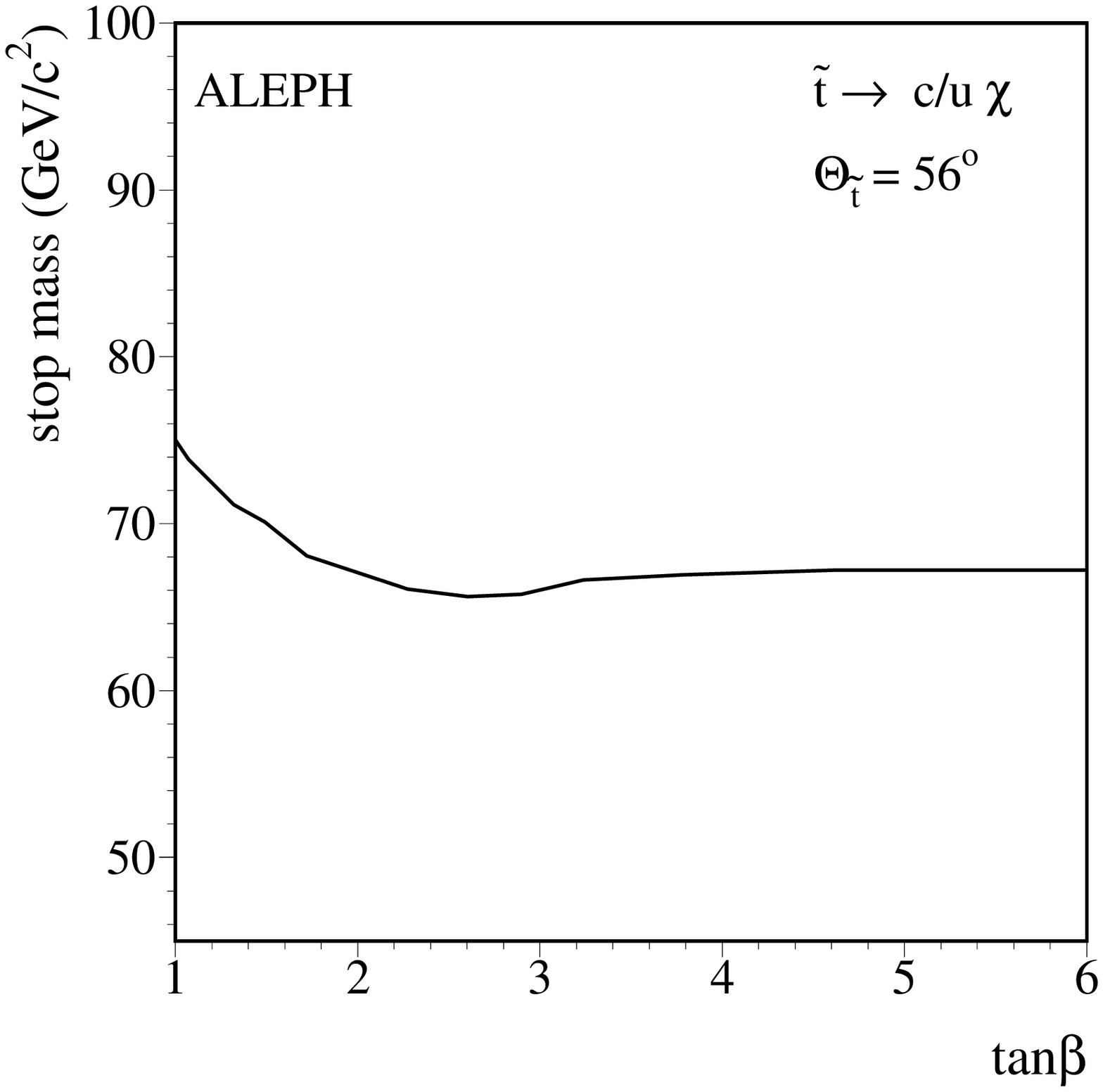,width=0.485\textwidth}
    \caption{(a) Excluded regions at 95\% C.L. 
      in the $M_{\neu}$ vs 
      $M_{\sto}$ plane from $\sto \to \charm / \u \neu$ searches;
      the excluded regions are given for $\thstop = 0^{\circ}$,
      corresponding to  maximum $\sto \sto$Z coupling, and for
      $\thstop = 56^{\circ}$, corresponding to vanishing
      $\sto \sto$Z coupling.
      The dark region in the small $\dM$ corridor is excluded 
      by the ``long-lived hadrons'' analysis.
      The CDF experiment result is also indicated.
      (b)~Stop mass lower limit at 95\% C.L. from the ``long-lived
      hadrons'' as a 
      function of $\log (c\tau_\sto /{\rm cm})$ for several $\dM$
      values, without any assumption on the relation between $\dM$ and
      the stop lifetime $\tau_\sto$.
      (c)~Stop mass lower limit at 95\% C.L. from the ``long-lived
      hadrons'' analysis as a function of $\tanb$, independent of the 
      other MSSM parameters.} 
    \label{stop}
  \end{center}
\end{figure}

The regions  excluded in the plane $(M_{\sto},M_\chi)$ under the
hypothesis of a dominant $\sto\to \charm/\u \, \chi$ decay
are shown in Fig.~\ref{stop}a for two values of the $\sto$ mixing
angle $\thstop$, $0^\circ$ and $56^\circ$, corresponding to maximal
and vanishing $\sto \sto$Z coupling, respectively. For
$8\GeVcc<\dM<M_{\W}+M_{\b}$, and using also CDF results~\cite{cdf},
the lower limit on $M_{\sto}$ is $92\GeVcc$, independent of $\thstop$.

The very small $\dM$ corridor is partially covered by the ``long-lived
hadrons'' analysis as indicated by the plain dark region in
Fig.~\ref{stop}a. The stop mass lower limit provided by the
``long-lived hadrons'' analysis is shown in Fig.~\ref{stop}b as a
function of $\log (c\tau_\sto/{\rm cm})$ for various $\dM$ values. The
smallest $\dM$ value considered is $1.6\GeVcc$, corresponding to the
``effective'' kinematic limit for the decay $\sto \to \charm
\chi$~\cite{llh}. Below that $\dM$ value, the stop decay mode is $\sto
\to \u \chi$, and the limit is $95\GeVcc$, given by the large lifetime
selection. The absolute mass lower limit obtained is $63\GeVcc$. It is
reached for $\dM=1.6\GeVcc$ and for a $c\tau_\sto$ value of $\sim
1\cm$. In that configuration of parameters, the ``AJ small $\dM$'' and
the ``long-lived hadrons intermediate lifetime'' selections are
combined.

\pagebreak

In the MSSM~\cite{mssm}, more restrictive constraints on the stop mass
can be obtained since $\dM$ and the stop lifetime are related. The
mass lower limit obtained by scanning over the relevant model
parameters as in Ref.~\cite{llh} is shown in Fig.~\ref{stop}c as a
function of $\tanb$. For any $\tan\beta$, the stop mass limit is
$65\GeVcc$, reached for $\tanb\sim2.7$.

\begin{figure}[p]
  \begin{center}
        \begin{picture}(0,0)
        \put(18,20){\mbox{\bf (a)}}
        \end{picture}
    \epsfig{file=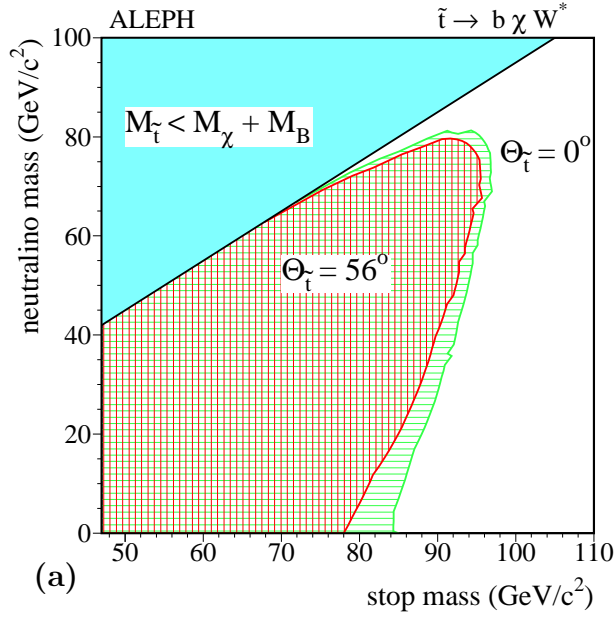,width=0.55\textwidth} \\
        \begin{picture}(0,0)
        \put(18,20){\mbox{\bf (b)}}
        \end{picture}
    \epsfig{file=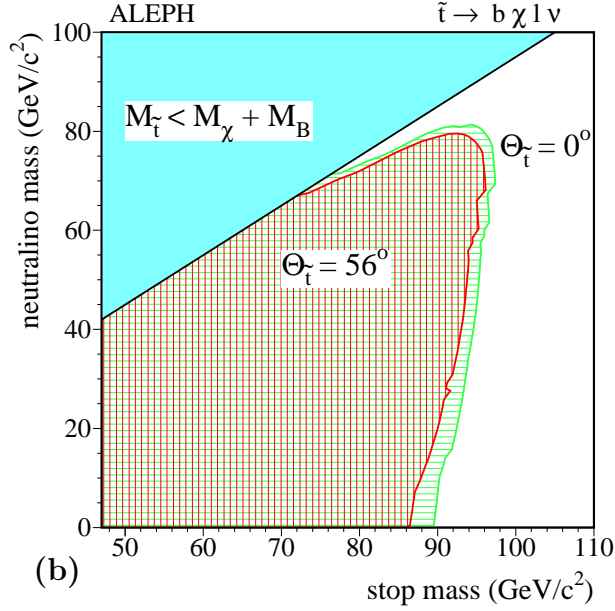,width=0.55\textwidth}
    \caption{Excluded regions at 95\% C.L. 
      in the $M_{\neu}$ vs 
      $M_{\sto}$ plane from $\fbody$ searches: (a)
      the W$^*$ modes or (b)
      the leptonic modes are assumed to be dominant for the 
      ${\rm f \bar{f}'}$
      final states. The excluded regions are given  
      for $\thstop = 0^{\circ}$,
      corresponding to  maximum $\sto \sto$Z coupling, and for
      $\thstop = 56^{\circ}$, corresponding to vanishing
      $\sto \sto$Z coupling.}
    \label{stop4b2}
  \end{center}
\end{figure}

Under the hypothesis that the decay $\fbody$ is dominant, the regions
excluded in the plane $(M_{\sto},M_\chi)$ are shown in
Fig.~\ref{stop4b2}a, for relative proportions of the possible ${\rm f
  \bar{f}'}$ final states as in W$^*$ decays. In Fig.~\ref{stop4b2}b
the leptonic modes $\b\chi\ell\nu$ (with equal branching ratios for
$\ell = \e$, $\mu$ and $\tau$) are assumed to be
dominant. The excluded regions are given for $\thstop = 0^\circ$ and
$\thstop = 56^\circ$. For $\dM>8\GeVcc$, the $\thstop$-independent
lower limits on $M_{\sto}$ are $78\GeVcc$ and $80\GeVcc$, for the two
cases of W$^*$ and leptonic final state dominance, respectively.

\begin{figure}[p]
  \begin{center}
        \begin{picture}(0,0)
        \put(18,20){\mbox{\bf (a)}}
        \end{picture}
    \epsfig{file=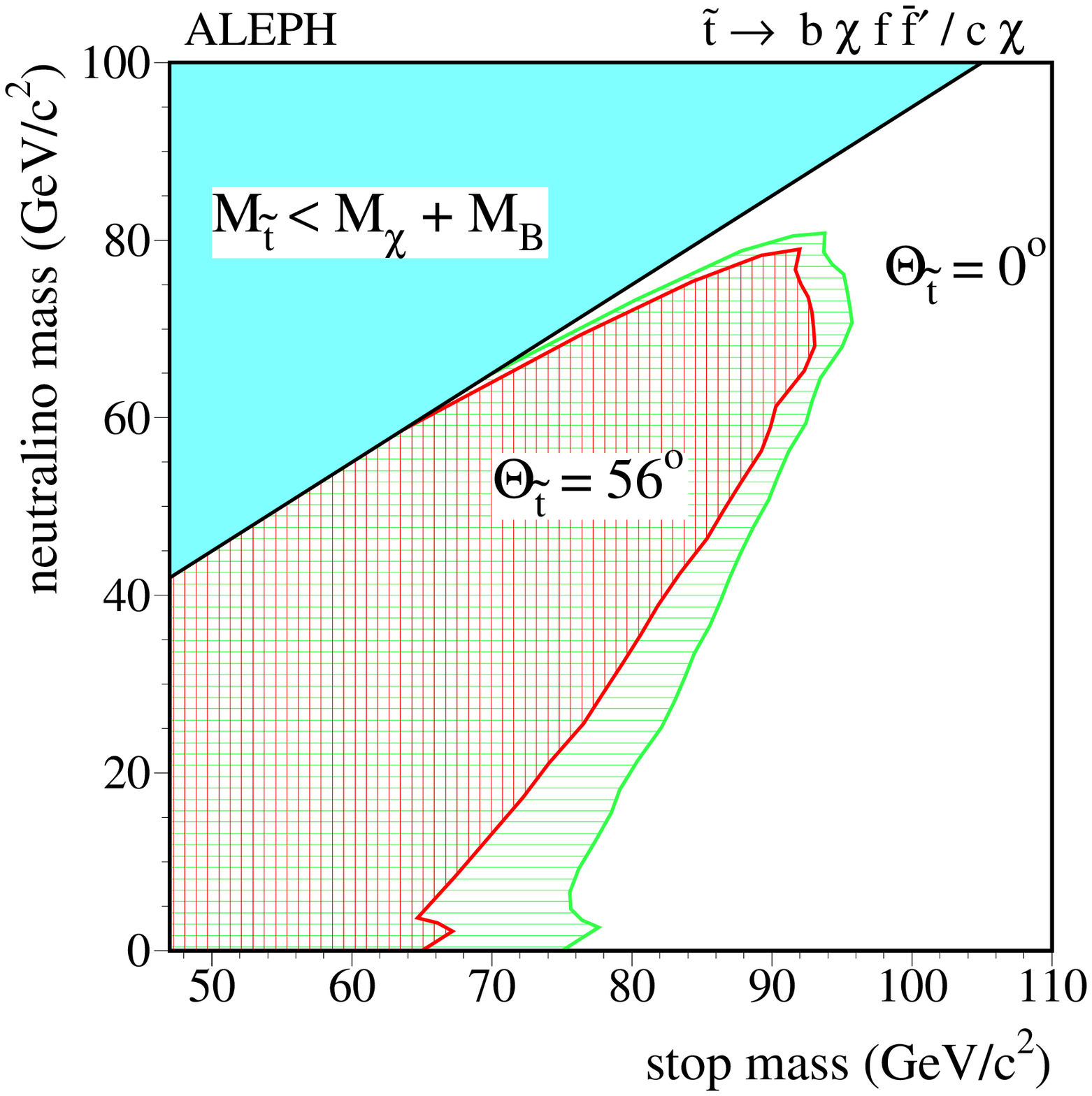,width=0.55\textwidth} \\
        \begin{picture}(0,0)
        \put(18,20){\mbox{\bf (b)}}
        \end{picture}
    \epsfig{file=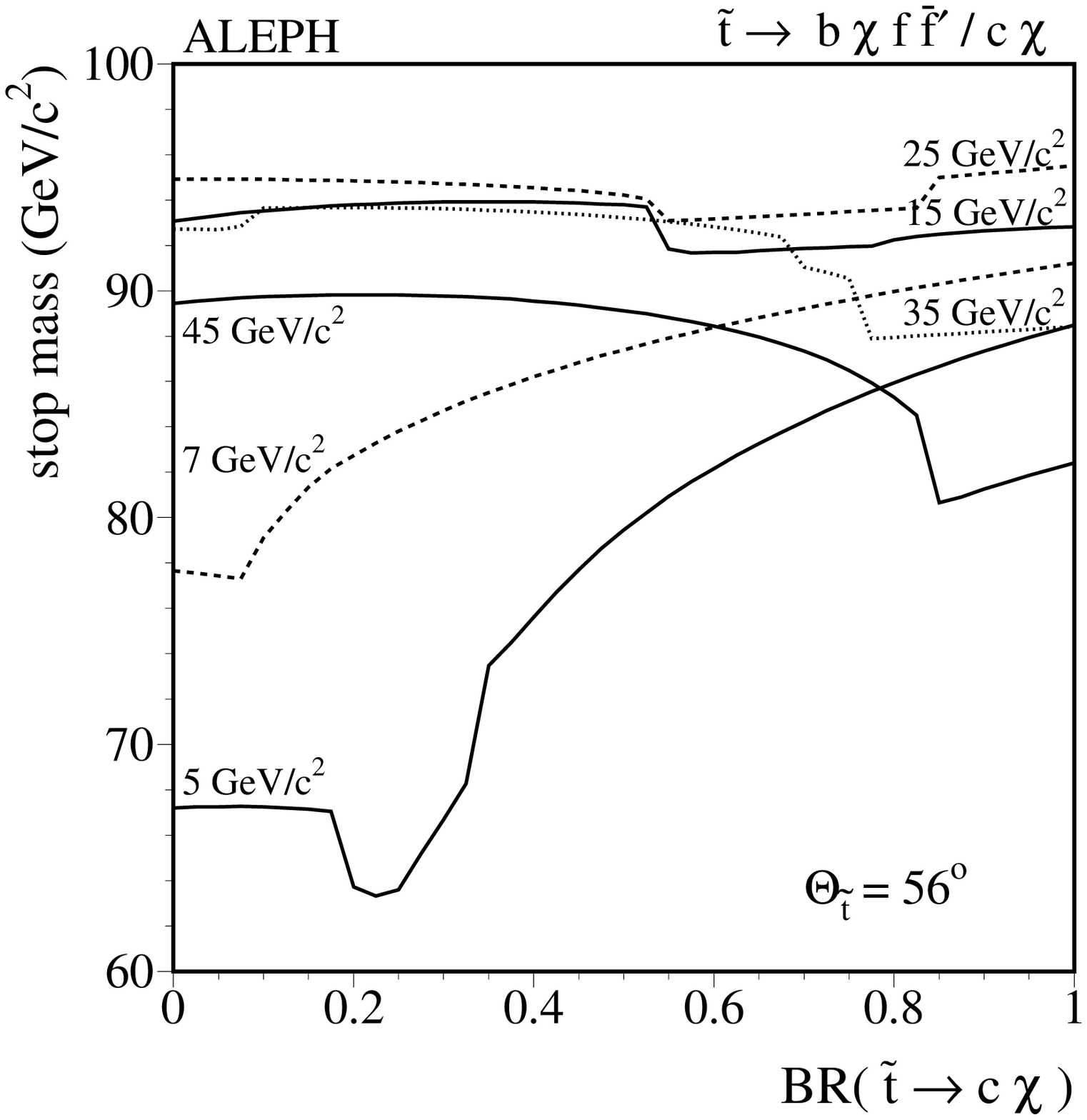,width=0.55\textwidth}
    \caption{(a) Branching ratio independent excluded regions at 95\% C.L. 
      in the $M_{\neu}$ vs 
      $M_{\sto}$ plane, from $\fbody$ and $\sto
      \to \charm \neu$ searches. The excluded regions are given  
      for $\thstop = 0^{\circ}$,
      corresponding to  maximum $\sto \sto$Z coupling, and for
      $\thstop = 56^{\circ}$, corresponding to vanishing
      $\sto \sto$Z coupling. (b) Limit on the stop mass at 95\%
      C.L. as a function of $\BR(\sto \to \charm \neu)$ for various $\dM$ values.
      The limits are given for $\thstop = 56^{\circ}$.}
    \label{stop4b1}
  \end{center}
\end{figure}

The combination of the AJ and MJ analyses allows constraints to be set
under the more general hypothesis that both the $\sto \to \charm \neu$
and $\fbody$ decay channels contribute to stop decays. The excluded
regions in the plane $(M_{\sto},M_\chi)$ are shown in
Fig.~\ref{stop4b1}a for $\thstop = 0^\circ$ and $\thstop =
56^\circ$. This result was obtained by arbitrarily varying the $\sto
\to \charm \neu$ branching ratio and the leptonic fraction in the
$\fbody$ decay, and by using the $\nnc$ prescription to determine the
appropriate combination of selections. The stop mass limit is shown in
Fig.~\ref{stop4b1}b as a function of the branching ratio $\BR(\sto \to 
\charm \neu)$ for several fixed $\dM$ values and for $\thstop = 56^\circ$. 
The smallest $\dM$ value considered is $5\GeVcc$, corresponding 
to the threshold for the production of a $\b$ quark in the final state. 
The lowest limit obtained is $63\GeVcc$; it is reached for $\dM=5\GeVcc$,
$\BR(\sto \to \charm\neu)=0.22$, and $\BR(\sto \to \b \neu \ell \nu)=0.55$.     

\begin{figure}[p]
  \begin{center}
        \begin{picture}(0,0)
        \put(18,20){\mbox{\bf (a)}}
        \end{picture}
    \epsfig{file=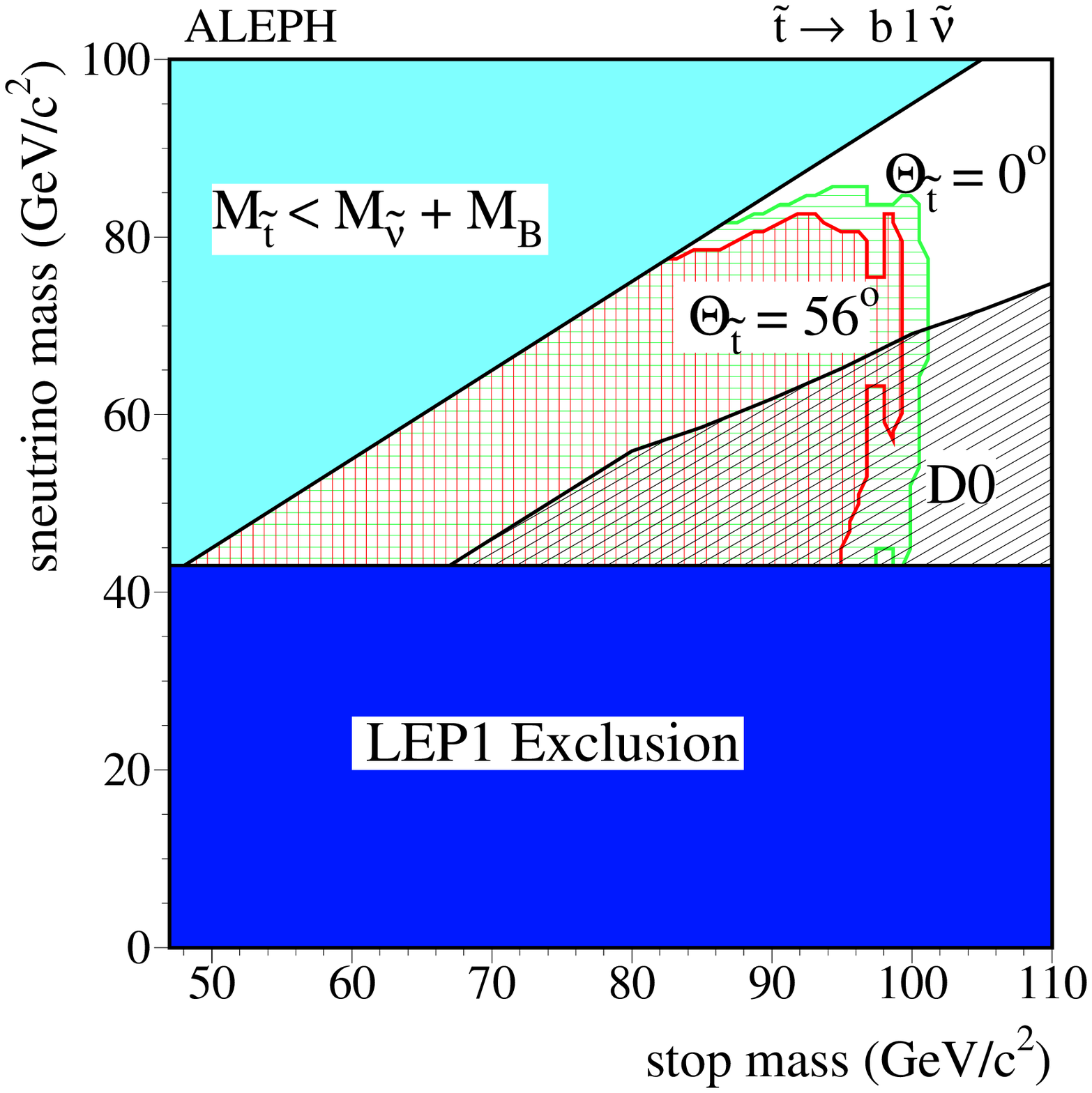,width=0.55\textwidth} \\
        \begin{picture}(0,0)
        \put(18,20){\mbox{\bf (b)}}
        \end{picture}
    \epsfig{file=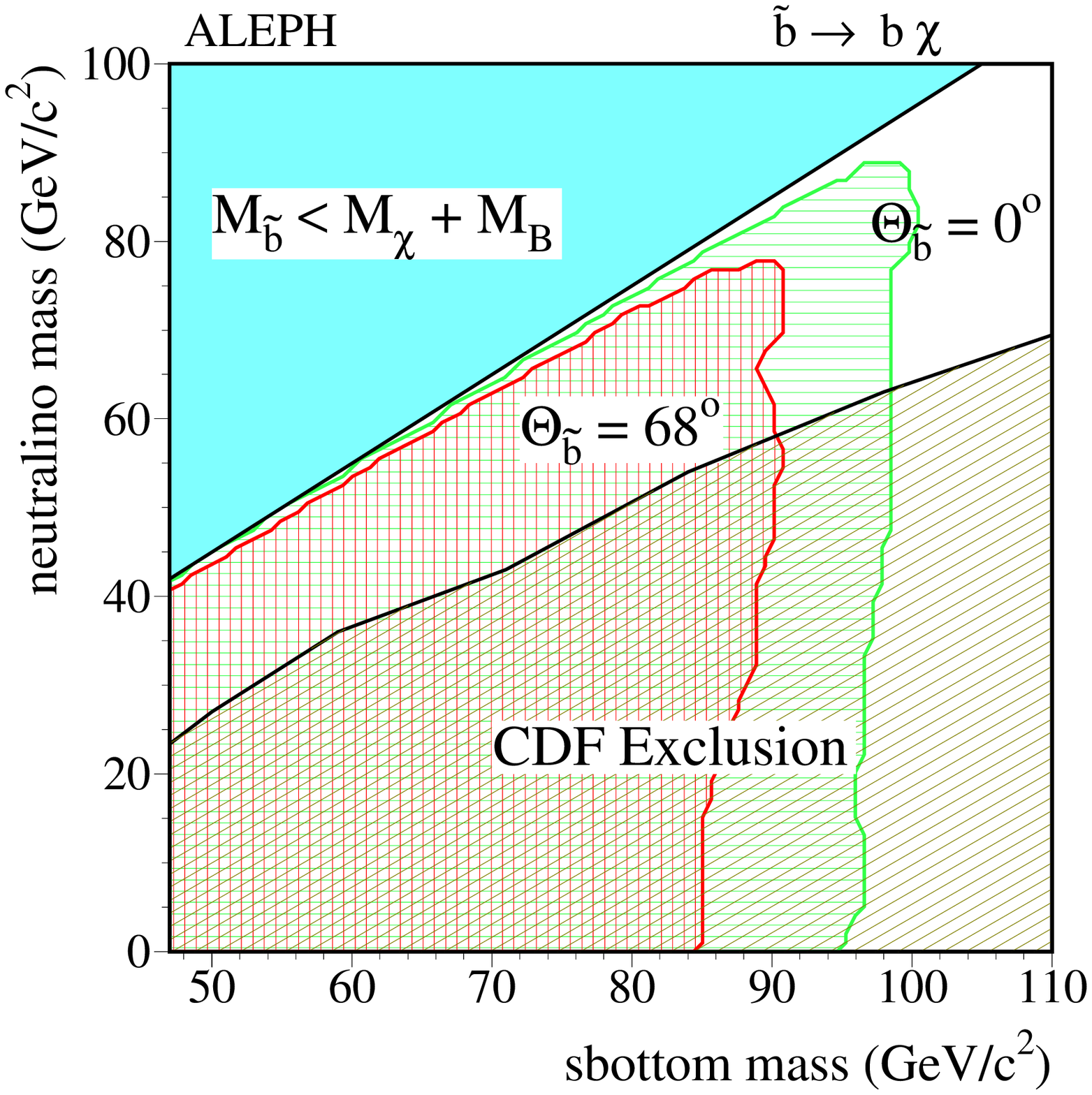,width=0.55\textwidth}
    \caption{(a) Excluded regions at  95\% C.L. in the 
      $M_{\snu}$ vs $M_{\sto}$ plane 
      from $\sto \to \b\ell\snu$ searches
      (equal branching fractions for the $\sto$ decay to $\e$,
      $\mu$, and $\tau$ are assumed). 
      The excluded regions are given for $\thstop = 0^{\circ}$,
      corresponding to  maximum $\sto \sto$Z coupling, and for
      $\thstop = 56^{\circ}$, corresponding to vanishing
      $\sto \sto$Z coupling. The regions excluded at LEP~1 and by the
      D0 experiment are also indicated.
      (b) Excluded regions at 95\% C.L. in the $M_{\neu}$ vs 
      $M_{\sbot}$ plane from $\sbot \to \b\neu$ searches. 
      The excluded regions 
      are given for $\thsbot =0^{\circ}$,
      corresponding to  maximum $\sbot \sbot$Z coupling, and for 
      $\thsbot = 68^{\circ}$, corresponding to vanishing $\sbot
      \sbot$Z coupling.
      The region excluded by the CDF experiment is also indicated. 
      }
    \label{stop2}
    \label{sbot}
  \end{center}
\end{figure}

Under the assumption that the $\sto \to \b\ell\snu$ decay mode is
dominant, with equal branching ratios for $\ell=\e$, $\mu$ and
$\tau$,  the excluded region in the plane $(M_{\sto},M_{\tilde{\nu}})$
is shown in Fig.~\ref{stop2}a. If $\dM >8 \GeVcc$, and using the LEP1
limit on the sneutrino mass and D0 results~\cite{d0}, the lower limit
on $M_{\sto}$ is $97\GeVcc$, independent of $\thstop$. The lower limit
is $82\GeVcc$ if the $\sto \to \b\tau\snu_\tau$ decay mode is dominant
and $\dM >8 \GeVcc$, independent of $\thstop$. 

The excluded region in the plane $(M_{\sbot},M_\chi)$ is shown in
Fig.~\ref{sbot}b under the assumption of a dominant $\sbot \to
\b\neu$ decay. Taking also the CDF exclusion~\cite{cdf} into account, a
lower limit of $89\GeVcc$ is set on $M_{\sbot}$, for any
$\sbot$ mixing angle and $\dM > 8\GeVcc$. The region excluded for
$\thsbot  = 0^{\circ}$, for which the $\sbot\sbot$Z coupling is maximal, is
also shown. 

\begin{figure}[p]
  \begin{center}
    \epsfig{file=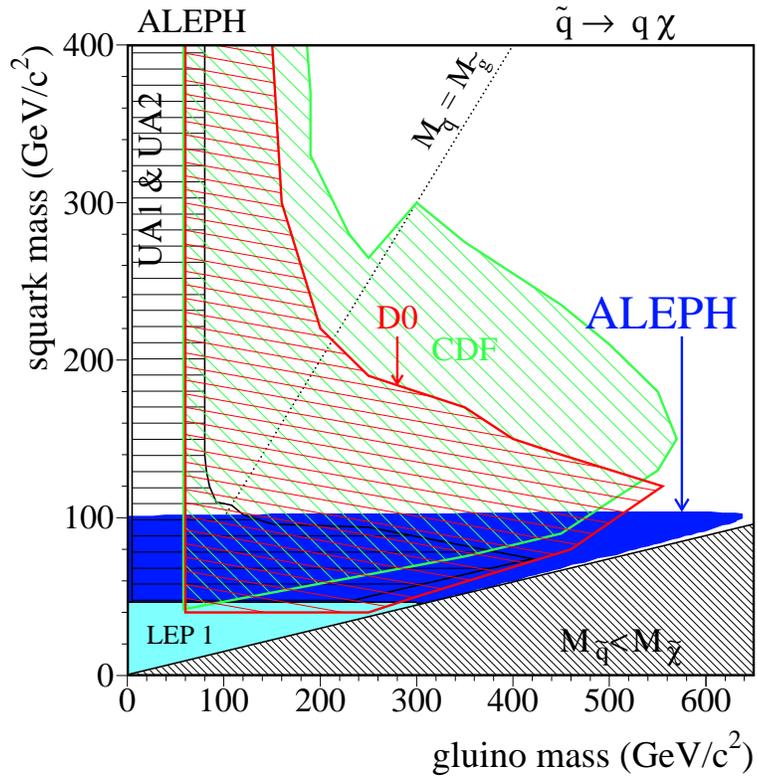,width=0.7\textwidth}
    \caption{Excluded regions at 95\% C.L. from the search for
      generic $\squa$ pairs, 
      assuming five mass-degenerate $\squa$  flavours.
      The results are shown 
      in the gluino-squark
      mass plane for $\tanb= 4$ and $\mu$ = $-400 \GeV$,
      together with results from experiments at $\p\bar{\p}$ colliders.}
    \label{dege}
  \end{center}
\end{figure}

As discussed in detail in Ref.~\cite{ALEPHsqtwo}, the results
of the search for acoplanar jets, with or without b tagging, can also
be translated into constraints on the mass of degenerate squarks. In
order to compare these results with those obtained at the
Tevatron~\cite{cdf, d0}, limits have been evaluated within the 
MSSM~\cite{mssm} under the following assumptions: a degenerate mass
$M_{\tilde{{\rm q}}}$ for all left-handed and right-handed
$\tilde{{\rm u}},\tilde{{\rm d}},\tilde{{\rm c}},\tilde{{\rm
    s}},\tilde{{\rm b}}$ squarks; lowest order GUT relation between
the soft supersymmetry breaking gaugino mass terms, allowing the
gluino and neutralino masses to be related; $\tanb\!=\!4$ and
$\mu\!=\!-400$~GeV. The results in the plane
$(M_{\mathrm{\glu}},M_{\squa})$ are shown in Fig.~\ref{dege}. Improved 
constraints are obtained in the region of small $\squa$ to $\chi$ mass
differences.

\pagebreak

\section{Conclusions}

Searches for signals of pair-produced scalar partners of quarks have been 
performed in the data sample of $207\ipb$ collected in the year 2000 
with the ALEPH detector at LEP, at centre-of-mass energies ranging from 
204 to $209\GeV$. The final state topologies studied arise from the decays 
$\sto \to \charm / \u \, \chi$, $\fbody$, $\sto \to \b \ell \snu$,
$\sbot \to \b \chi$, and $\squa \to \q \chi$. The four-body stop decay channel 
was analysed for the first time at LEP, and the corresponding selections were
extended to the $675\ipb$ of data collected by ALEPH at centre-of-mass 
energies of $183\GeV$ and above. All numbers of candidate events observed are 
consistent with the backgrounds expected from Standard Model processes. 
The results of these searches, combined with earlier ones obtained with data 
collected from 1997 to 1999, have been translated into improved mass lower 
limits, of which relevant examples are given in Table~\ref{tab:lim}. 
In particular, a 95\% C.L. lower limit of $63\GeVcc$ has been set on the 
stop mass, irrespective of its lifetime and decay branching ratios.

\begin{table}[htbp]
  \caption{\small Lower limits on stop and sbottom masses
    in some relevant cases. All limits are valid for any value of the
    mixing angle.  
    \label{tab:lim}}
  \begin{center}
    \begin{tabular}{|c|c|c|l|l|}
      \hline
             & 95\% C.L.  & $\dM$      &  Dominant                         &          \\
             & Mass Limit & range      &  decay                            &          \\
      Squark & [$\GeVcc$] & [$\GeVcc$] &  channel(s)                       & Comments \\ \hline
\rule{-4pt}{14pt}
      $\sto$ & 92         & $>8$       &  $\sto \to \charm \chi$           & CDF result~\cite{cdf} used \\
             & 78         & $>8$       &  $\sto \to \b \chi \W^*$          &          \\
             & 97         & $>8$       &  $\sto \to \b \ell \snu$
             & LEP~1, D0 result~\cite{d0} used\\
             & 63         & any        &  any & any branching
             ratios, any lifetime\\
\hline
\rule{-4pt}{14pt}
     $\sbot$ & 89         & $>8$       &  $\sbot \to \b \chi$              & CDF result~\cite{cdf} used \\
      \hline
    \end{tabular}
  \end{center}
\end{table}

\section*{Acknowledgements}
We wish to congratulate our colleagues from the accelerator
divisions for the successful operation of LEP at high energies. 
We would also like to express our gratitude to the engineers
and support people at our home institutes without whom this work
would not have been possible. Those of us from non-member states
wish to thank CERN for its hospitality and support.


\newpage

\newpage

\newpage

\end{document}